%% file: ex1489c.tex
\documentclass[twocolumn,floatfix]{revtex4-2}
\pdfoutput=1
\usepackage{graphicx}
\usepackage{dcolumn}
\usepackage{longtable}
\usepackage{amsmath}
\usepackage{amssymb}
\usepackage{color}
\usepackage{isotope}

\begin{document}
\title{Parameter-free deformation variables of the proxy-SU(3) symmetry in even-even actinide, superheavy and hyperheavy nuclei with ${\bf Z=82-126}$,
 ${\bf N=82-258}$}

\author
{Dennis Bonatsos$^1$, V. K. B. Kota$^2$, Andriana Martinou$^1$,  S. K. Peroulis$^1$, D.~Petrellis$^3$, P. Vasileiou$^4$, T. J. Mertzimekis$^5$, and N. Minkov$^6$ }

\affiliation
{$^1$Institute of Nuclear and Particle Physics, National Centre for Scientific Research ``Demokritos'', GR-15310 Aghia Paraskevi, Attiki, Greece}

\affiliation
{$^2$ Physical Research Laboratory, Ahmedabad 380 009, India} 

\affiliation
{$^3$ Physics Department, Aristotle University of Thessaloniki, Thessaloniki GR-54124, Greece}

\affiliation
{$^4$ Horia Hulubei National Institute for R{\&}D in Physics and Nuclear Engineering, Strada Reactorului 30, POB MG6, RO-077125
Bucharest-M\v{a}gurele, Romania}

\affiliation
{$^5$  Department of Physics, National and Kapodistrian University of Athens, Zografou Campus, GR-15784 Athens, Greece}

\affiliation
{$^6$Institute of Nuclear Research and Nuclear Energy, Bulgarian Academy of Sciences, 72 Tzarigrad Road, 1784 Sofia, Bulgaria}

\begin{abstract}

Superheavy and hyperheavy nuclei are one of the frontiers of nuclear structure nowadays, while also for many actinides rather limited experimental information exists.
Therefore, theoretical methods providing parameter-independent predictions for these nuclei are of particular interest. Such a method is the proxy-SU(3) approximation to the shell model, which has been adequately tested against experimental data in medium-mass and heavy nuclei up to the rare earth region, and has been found to provide reliable, parameter-independent predictions for the collective deformation variables $\beta$ and $\gamma$. Within the proxy-SU(3) approach, the SU(3) symmetry of the 3-dimensional harmonic oscillator, which is destroyed beyond the $sd$ shell by the strong spin-orbit interaction, is restored through a unitary transformation. For each nucleus, the most symmetric irreducible representation (irrep) allowed by the Pauli principle and the short-range nature of the nucleon-nucleon interaction, called the highest-weight (hw ) irrep in mathematical language, 
is found to suffice, except in cases in which the hw irrep turns out to be completely symmetric, so that the next highest weight (nhw) irrep has also to be included. 
In this article we provide a full collection of the hw and nhw irreps, as well as of the corresponding parameter-free predictions for the deformation variables $\beta$ and $\gamma$,
for all atomic nuclei ranging from $Z=82$, $N=82$ to $Z=126$, $N=258$. Several cases exemplifying the use of the collected results for studying the prolate to oblate shape transition, mirror symmetries, as well as the evolution of the collective variables along the valley of stability are also considered. 
 
\end{abstract}

\maketitle

\section{Introduction}

One of the frontiers of nuclear structure nowadays is the production of superheavy and hyperheavy nuclei, as well as the theoretical prediction of their properties, of which the experimental study is a very hard task. The term superheavy nuclei \cite{Seaborg1991,Hoffman2000} is used for transactinides with $104 \leq Z \leq 126$ (see Refs. \cite{Hofmann1998,Hofmann2000,Erler2012,Oganessian2015,Oganessian2017,Nazarewicz2018,Giuliani2019} for relevant reviews), while for nuclei beyond $Z=126$ the term hyperheavy nuclei is used \cite{Decharge1999,Bender2001}. A variety of theoretical approaches has been used for the study of superheavy and hyperheavy nuclei, including non-relativistic mean-field calculations with the Skyrme \cite{Bender1998}, Gogny \cite{Egido2020}, and Giessen \cite{Malov2021} interactions, relativistic mean-field calculations \cite{Lalazissis1996,Patra1999,Patra2000,Mehta2015}, 
relativistic energy density functionals \cite{Prassa2012,Abusara2012,Prassa2013,Agbemava2015,Raeder2018,Afanasjev2018,Agbemava2019,Afanasjev2020,Agbemava2021}, as well as the collective model with parameters adjusted through microscopic symmetry considerations \cite{Ermamatov2016}, while their synthesis within a neutron-star environment has also been proposed \cite{Veselsky2022}. 

A disadvantage of these methods is that they contain free parameters, which are fixed to the data regarding lighter nuclei, with their use arbitrarily extrapolated to heavier regions, where there are practically no data to compare the theoretical predictions against, although the actinides, having $84\leq Z \leq 102$, can serve as a reasonable test-ground for such theories, since for the actinides, especially for the ligher ones, a substantial body of nuclear data exist \cite{ensdf}. 

It would be certainly desirable to have a theory which could provide reliable predictions for superheavy and hyperheavy nuclei. The proxy-SU(3) scheme \cite{Bonatsos2017a,Bonatsos2017b,Bonatsos2023}  could be  a useful tool in this direction, since it is based on the SU(3) symmetry of the 3-dimensional harmonic oscillator (3D-HO) \cite{Wybourne1974,Moshinsky1996,Iachello2006}, which is restored beyond the nuclear $sd$ shell through a unitary transformation \cite{Martinou2020}, as well as on the short-range nature of the nucleon-nucleon interaction \cite{Martinou2021b}. The proxy-SU(3) scheme 
is known to produce reliable parameter-independent predictions for the $\beta$ and $\gamma$ collective variables up to the rare earth region \cite{Bonatsos2017b,Sarantopoulou2017,Bonatsos2023,Bonatsos2026}. It would be interesting to extend its application to the actinides, which can serve as a test-ground for this extension, as well as to the superheavy and hyperheavy regions. 

In order to carry out this task, one needs to determine in these heavy regions the most symmetric irreducible representations (irreps) of SU(3) allowed by the Pauli principle, called the highest weight irreducible representations (hw irreps) in mathematical language \cite{Martinou2021b}. The physical meaning of the hw irreps, as well as the way of their determination, have been discussed in detail in Refs. \cite{Martinou2018,Martinou2021b}. Although in principle there are codes able of performing this task \cite{Draayer1989a,Langr2019,Alex2011}, the computational load is non-negligible. However, these calculations have to be carried out once and for all. After tabulating the relevant irreps, one can easily use them for many purposes. 

It is the purpose of the present work to provide full tables of hw irreps for the actinide, superheavy and hyperheavy regions. In addition to the hw irreps, the next hw irreps (nhw irreps) will also be tabulated, since they become necessary \cite{Bonatsos2024b,Bonatsos2025} in the rare case in which the hw irrep turns out to be fully symmetric, something not expected for a system of many fermions. The proxy-SU(3) parameter-independent predictions for the collective $\beta$ and $\gamma$ variables will also be tabulated, 
with the aim of providing a concrete set of predictions in regions in which scarce experimental information exists.

In Section II of the present study a brief review of the proxy-SU(3) approximation is given, while in Section III the way in which the hw and nhw irreps are determined is described. In Section IV the calculation of the collective variables $\beta$ and $\gamma$ is described, while a few examples in which these results can be used are given in Section V. Finally,  Section VI contains a summary of the present results and a brief outlook for further work. 

\section{The proxy-SU(3) approximation to the shell model} 

In its simplest version the spherical nuclear shell model \cite{Mayer1948,Mayer1949,Haxel1949,Mayer1955,Heyde1990,Talmi1993} is based on a 3-dimensional harmonic oscillator (3D-HO) \cite{Wybourne1974,Moshinsky1996,Iachello2006}, to which the spin-orbit force is added, in order to reproduce the nuclear magic numbers 2, 8, 20, 28, 50, 82, 126, \dots. The shells of the 3-dimensional harmonic oscillator, which exhibits magic numbers at 2, 8, 20, 40, 70, 112, 168, \dots, own a U(${\cal N}$) symmetry with ${\cal N}=n(n+1)/2$, where $n$ is the order of the shell, which is known in all cases to possesss 
an SU(3) subalgebra \cite{Bonatsos1986}. For example, the $sd$ shell, which has $n=3$, is characterized by the U(6) overall symmetry. 

The spin-orbit interaction destroys the SU(3) symmetry beyond the $sd$ shell, since it acts most drastically to the levels with the highest $j$ in each shell, pushing them down to the shell below. Thus each shell is losing a set of orbitals, called the deserting orbitals, to the shell below, while in parallel it gains a different set of orbitals, called the intruder orbitals, from the shell above. The original orbitals remaining in the shell are called the normal parity orbitals. For example, the $pfh$ shell, comprised of the $3p_{1/2}$, $3p_{3/2}$, $2f_{5/2}$, $2f_{7/2}$, $1h_{9/2}$, $1h_{11/2}$ orbitals, loses the $1h_{11/2}$ orbitals to the $sdg$ shell lying below it, while in parallel it is gaining the $1i_{13/2}$ from the shell above, which is the $sdgi$ shell. The proxy-SU(3) approximation \cite{Bonatsos2017a,Bonatsos2017b,Bonatsos2023} is based on the fact that 
the deserting orbitals and the intruder orbitals (except the one with the highest angular momentum projection $m$, which lies highest in energy) are very similar, allowing the intruder set (without its highest-lying member) to act as a ``proxy'' of the deserting set. 

The quality of this approximation has been first tested within the Nilsson model \cite{Nilsson1955,Nilsson1995}, an axially deformed version of the spherical shell model, in which the orbitals are characterized by the quantum numbers $K[N n_z \Lambda]$, where $N$ is the number of oscillator quanta, $n_z$ is the number of oscillator quanta along the axis of cylindrical symmetry, while $K$ and $\Lambda$ are the projections of the total angular momentum  and the orbital angular momentum along the axis of cylindrical symmetry. It has been seen \cite{Bonatsos2017b} that the replacement of the intruder orbitals (except the highest-lying one) by the deserting orbitals has minimal influence on the numerical values of the single-particle energies appearing in the Nilsson diagrams, thus supporting the idea that the deserting orbitals can act as ``proxies'' of the intruder orbitals (except the highest-lying one), thus restoring the SU(3) symmetry of the shell. 
It has also been seen \cite{Bonatsos2017b} that the larger the shell, the better the approximation becomes, which means that the proxy-SU(3) symmetry should be particularly appropriate for the actinide, superheavy and hyperheavy nuclei. 

From the physics point of view, one can observe that the intruder orbitals are characterized by Nilsson quantum numbers higher by 0[110] in comparison to the Nilsson quantum numbers of the deserting orbitals. We shall call such pairs of orbitals the 0[110] pairs. It is clear that orbitals forming such a pair
are characterized by identical values of all angular momentum projections, differing only by one oscillator quantum along the direction of the axis of cylindrical symmetry, thus having very similar and largely overlapping shapes \cite{Bonatsos2013}. Actually the idea of 0[110] pairs first emerged from the observation that for 0[110] pairs the proton-neutron interaction is maximized. This observation has been made in deformed rare earths through the study of double differences of binding energies, acting as a filter isolating the valence proton-neutron interaction, which is maximized between proton and neutron orbitals possessing similar shapes \cite{Cakirli2005,Cakirli2006,Stoitsov2007,Cakirli2010,Cakirli2013,Cakirli2025}. 

From the mathematical point of view, the proxy-SU(3) symmetry has been connected to the shell model \cite{Martinou2020,Bonatsos2020b}, through the realization that the intruder orbitals and their ``proxies'' are connected through a unitary transformation \cite{Martinou2020}. A similar situation had been found earlier in the framework of the pseudo-SU(3) symmetry \cite{Arima1969,Hecht1969,RatnaRaju1973,Draayer1982,Draayer1983,Draayer1984,Bahri1992,Ginocchio1997}, which is a different approximation to the shell model taking advantage of the restoration of the 3D-HO SU(3) symmetry. Within the pseudo-SU(3) scheme, the 3D-HO SU(3) symmetry is restored only for the normal parity orbitals, i.e. the orbitals remaining in a shell after the desertion of the orbitals with the highest value of the total angular momentum \cite{RatnaRaju1973}. The restoration of  the SU(3) symmetry is achieved by mapping the normal parity orbitals onto the orbitals of the shell below, a process equivalent to a unitary transformation \cite{Castanos1992a,Castanos1992b,Castanos1994}, albeit a unitary transformation more complicated than the one appearing in the proxy-SU(3) case \cite{Martinou2020}. The main difference between the two approaches regards the intruder orbitals, which in proxy-SU(3) are taken into account in the SU(3) approximation, while in pseudo-SU(3) they remain alone, having to be treated separately by shell-model techniques \cite{Draayer1983,Draayer1984}. It is an intriguing fact that both approximations give similar numerical results \cite{Bonatsos2020a} when involved in calculations of the collective variables $\beta$ and $\gamma$, provided that the appropriate hw irrep, to be discussed in the next section, is used in both frameworks. 

It should be remarked that the ``proxy'' orbitals in proxy-SU(3) have parity opposite to the parity of the orbitals they are replacing. Similarly, in pseudo-SU(3) the orbitals of the lower shell replacing the normal parity orbitals possess the opposite parity. The parity issue makes no difference in the case of even-even nuclei, in which only pairs of orbitals of the same parity are used. The extension of the proxy-SU(3) symmetry to odd nuclei, which has not been carried out yet, might require parity projection techniques \cite{Ring1980}.

Another point on which further work will be needed, is the inclusion of the pairing interaction \cite{Ring1980,Brink2005}, which is known to break the SU(3) symmetry, in the proxy-SU(3) scheme, probably following a path analog to the one used \cite{Troltenier1995a,Troltenier1995b,Troltenier1996} for taking into account the pairing interaction within the pseudo-SU(3) framework.

\section{Highest weight irreducible representations} \label{irreps} 

Once the SU(3) symmetry has been restored, one has to choose the appropriate irreducible representations of SU(3) characterizing the valence protons and the valence neutrons of each specific nucleus. This selection is based on determining the most symmetric SU(3) irrep allowed by the Pauli principle \cite{Martinou2021b}. 
Because of the short-range nature of the nucleon-nucleon interaction, this irrep will be lying lowest in energy. From the mathematical point of view, this turns out to be the highest weight irrep. It should be remarked that the highest weight irrep is identical to the irrep exhibiting the highest eigenvalue of the second order Casimir operator of SU(3), $C_2$, up to the midshell, but this identity is broken in the upper half of the shell \cite{Bonatsos2017b,Martinou2021b}. The Elliott notation $(\lambda, \mu)$ \cite{Elliott1958a,Harvey1968} will be used for the SU(3) irreps.

The hw irreps and nhw irreps for the various shells, calculated using the code of Ref. \cite{Draayer1989a} and corroborated through a simple formula given in Ref. \cite{Kota2018}, are shown in Table  I 
(see also Appendix A for the description of some regularities appearing in the table). From this table one can then determine the hw irrep corresponding to the valence protons of the given nucleus, $(\lambda_p, \mu_p)$, as well as  the hw irrep corresponding to the valence neutrons of the given nucleus, $(\lambda_n, \mu_n)$. Then the hw irrep of the whole nucleus will be given by the most stretched irrep \cite{Draayer1983,Draayer1984} coming from the combination of the hw proton irrep and the hw neutron irrep, given by 
\begin{equation}
(\lambda, \mu) =  (\lambda_p, \mu_p) + (\lambda_n, \mu_n) = (\lambda_p+\lambda_n, \mu_p+\mu_n). 
\end{equation}
For most nuclei, the hw irrep turns out to have $\mu\geq 4$, which means that it can accommodate the ground state band (lowest band with $K=0$) the quasi-$\gamma_1$ band (lowest band with $K=2$), as well as the lowest $K=4$ band \cite{Elliott1958a,Harvey1968}. This is a desirable result, since the gsb and the quasi-$\gamma_1$ band are known to be connected by relatively strong interband $B(E2)$ transition rates \cite{ensdf,Bonatsos2025} and to bear several structural similarities \cite{Jolos2006,Jolos2007,Minkov1997,Minkov1999,Minkov2000,Bonatsos2021}. This would have been impossible if they were belonging to different irreps, since $B(E2)$ transitions connecting different irreps are not allowed. The next low-lying irrep, called the next hw irrep (nhw irrep) will contain the first excited $K=0$ band, also called the quasi-$\beta_1$ band, as well as the second $K=2$ band, also called the quasi-$\gamma_2$ band, and so on. 

At this point one has to be careful with the nature of the experimentally known first excited $0^+$ states \cite{Garrett2001,Aprahamian2025} and the bands built on them. First of all, not all $0^+$ states correspond to collective excitations \cite{Garrett2001,Aprahamian2025}. In addition, the interpretation of the $\beta$ bands as collective vibrations has been recently extensively questioned \cite{Sharpey2008,Sharpey2010,Sharpey2011a,Sharpey2011b,Sharpey2019}. 

In any case, it is desirable to know the nhw irrep for each nucleus, not only because the first excited $K=0$ band will lie in it, but also 
in the rare special cases in which the hw irrep of a nucleus turns out to have $\mu=0$, meaning that it can accommodate the gsb alone, while the lowest lying $K=2$ band (the quasi-$\gamma_1$ band) will  have to be pushed into the nhw irrep. As remarked above, this would be quite unphysical, since the gsb and the $\gamma_1$ bands carry several structural similarities \cite{Jolos2006,Jolos2007,Minkov1997,Minkov1999,Minkov2000,Bonatsos2021} and are connected by substantial B(E2) interband transitions \cite{ensdf,Bonatsos2025}. From the theoretical point of view, it would also be quite unexpected to have a system of many fermions exhibiting a completely symmetric hw irrep.   

The question then comes on how to determine the nhw irrep for each nucleus. While the full calculation of all irreps produced by combining the available proton irreps and neutron irreps is possible for light nuclei \cite{Kota2024,Kota2025}, this task becomes impossible in heavy nuclei. Therefore, a plausible shortcut is used. For each nucleus the proton hw and nhw irreps are coupled in all possible ways to the neutron hw and nhw irreps, and then the hw and nhw irreps are selected among the possible results by following the rules of Ref. \cite{Hecht1965}, namely that among the various SU(3) irreps the one with the highest value of $2\lambda+\mu$ should be chosen, while among irreps possessing the same value of $2\lambda+\mu$, the irrep with the highest value of $\mu$ should be chosen.

Relevant tables for the hw and nhw irreps for actinide, superheavy and hyperheavy nuclei with $Z=82$-126 and $N=82$-258 are given in Appendix B. 
The following observations can be made. 

In cases in which both the proton hw irrep and the neutron hw irrep have $\mu \neq 0$, if $(\lambda,\mu)$ is the hw irrep, then the nhw irrep turns out to be $(\lambda+2,\mu-4)$. 

From Table I it is seen that $\mu=0$ occurs at the valence particle numbers $M=2$, 6, 12, 20, 30, 42, 56, while $\mu=2$ occurs only for $M=4$. 
In the tables of Appendix B we see that when the valence particle numbers $M=2$, 4, 6, 12, 20, 30, 42, 56 occur for both valence protons and valence neutrons, the nhw following the hw irrep $(\lambda,\mu)$ is different from $(\lambda+2,\mu-4)$. If only one of the numbers of the valence protons or the valence neutrons belongs to the list, the rule that the hw irrep $(\lambda,\mu)$ is followed by a nhw irrep of the form $(\lambda+2,\mu-4)$ is preserved.  

These three different cases will be clarified through the following three examples.

Let us first consider the nucleus \isotope[234][92]{U}$_{142}$.  One has to consider the hw and nhw irreps for the 10 valence protons in U(21), which from Table I are $P_1$=(30,4) and $P_2$=(32,0) respectively, as well as the hw and nhw irreps for the 16 valence neutrons in U(28), which are $N_1$=(50,8) and $N_2$=(52,4) respectively. Neither of the numbers of the valence protons and neutrons, 10 and 16 respectively, belongs to the set $M=2$, 4, 6, 12, 20, 30, 42, 56. Obviously there are 4 possible combinations, $P_1+N_1=(80,12)$, $P_1+N_2=(82,8)$,  $P_2+N_1=(82,8)$, and  $P_2+N_2=(84,4)$. The rule for selecting the irrep with the highest weight among them, is given in Ref. \cite{Hecht1965}. The highest weight irrep $(\lambda,\mu)$ is characterized by the highest value of $2\lambda+\mu$, while among irreps with equal values of $2\lambda+\mu$, the irrep with the highest $\mu$ wins. In the present case, all irreps have $2\lambda+\mu=172$, therefore (80,12) is the hw irrep, while (82,8) is the nhw irrep. Notice that the nhw irrep is given by $(\lambda+2,\mu-4)$ if the hw irrep is $(\lambda,\mu)$.

Let us next consider the nucleus \isotope[236][94]{Pu}$_{142}$. From Table I we see that the hw and nhw irreps for the 12 valence protons in U(21) are $P_1$=(36,0) and $P_2$=(28,10),
while for the 16 valence neutrons in U(28) the hw and nhw irreps are $N_1$=(50,8) and $N_2$=(52,4) respectively, as above.
We remark that the 12 valence protons coincide with one of the numbers in the list of  $M=2$, 4, 6, 12, 20, 30, 42, 56.
 The 4 possible combinations are $P_1+N_1=(86,8)$, $P_1+N_2=(88,4)$,  $P_2+N_1=(78,18)$, and  $P_2+N_2=(80,14)$. The rule for selecting the irrep with the highest weight among them \cite{Hecht1965} has been mentioned above. In the present case, the irreps (86,8) and (88,4) have $2\lambda+\mu=180$, while the irreps (78,18) and (80,14) have $2\lambda+\mu=174$.
Therefore the first two irreps have priority for being the hw irrep, since they have the highest value of $2\lambda+\mu$. Among them, (86,8) is the hw irrep,  since it has higher $\mu$ than the irrep (88,4), while the irrep (88,4) is the nhw irrep. We remark that the rule of the hw irrep $(\lambda,\mu)$ followed by the nhw irrep $(\lambda+2,\mu-4)$ is restored.  

As a third example, let us consider the nucleus \isotope[240][94]{Pu}$_{146}$, for which both the 12 valence protons and the 20 valence neutrons belong to the $M=2$, 4, 6, 12, 20, 30, 42, 56 list. For the 12 valence protons in the 82-126 shell we see in the U(21) columns of Table I that the hw and nhw irreps are $P_1=(36,0)$ and $P_2=(28,10)$ as above, while for the 20 valence neutrons in the 126-184 shell we see in the U(28) columns of Table I that the hw and nhw irreps are $N_1=(60,0)$ and $N_2=(50,14)$. The 4 possible combinations are then $P_1+N_1=(96,0)$, $P_1+N_2=(86,14)$,  $P_2+N_1=(88,10)$, and  $P_2+N_2=(78,24)$. We remark that the quantity $2\lambda+\mu$ for these 4 irreps obtains the values 192, 186, 186, 180 respectively. Therefore (96,0), which possesses the highest $2\lambda+\mu$ value, is the hw irrep, according to the above mentioned rules of Ref. \cite{Hecht1965}, while the irreps (86,14) and (88,10), which have the same $2\lambda+\mu$ value of 186, compete for the nhw irrep place, the winner, according to the rules of Ref. \cite{Hecht1965}, being (86,14), since it possesses the highest $\mu$ value between them. We see that the hw irrep (96,0) and the nhw irrep (86,14) do not follow the rule that the nhw is given by $(\lambda+2,\mu-4)$ if the hw irrep is $(\lambda,\mu)$.

\section{The collective variables $\beta$ and $\gamma$} \label{coll}

The SU(3) symmetry \cite{Kota2020} has played a key role in building a bridge between the microscopic nuclear shell model and the collective model of Bohr and Mottelson \cite{Bohr1952,Bohr1953,Bohr1998a,Bohr1998b}, in which the nuclear shape is described in terms of the variables $\beta$, expressing the deviation from sphericity towards axially symmetric quadrupole deformation, and $\gamma$, regarding the deviation from axial symmetry towards triaxiality. This has been achieved by Elliott in 1958 \cite{Elliott1958a,Elliott1958b,Elliott1963,Elliott1968,Harvey1968}, by showing that within the $sd$ shell of the spherical shell model, intrinsic states exhibiting quadrupole deformation can be determined though the SU(3) classification scheme. The group theoretical framework of this finding is the SU(3) subalgebra of the U(6) algebra characterizing the $sd$ shell. It may be noticed at this point that Elliott also provided an alternative classification \cite{Elliott1958a} within the $sd$ shell, in terms of the O(6) subalgebra of U(6). These classifications have been used much later in the framework of the SU(3) \cite{Arima1978} and O(6) \cite{Arima1979} dynamical symmetries of the Interacting Boson Model \cite{Arima1975,Arima1976,Iachello1987}, in which medium mass and heavy nuclei are approximated by bosons corresponding to their valence proton pairs and valence neutron pairs counted from the nearest closed shell.   

The invariants of the collective model of Bohr and Mottelson are $\beta^2$ and $\beta^3 \cos 3\gamma$ \cite{Bohr1998b}, while the invariants of SU(3) are the second- and third-order Casimir operators, $C_2$ and $C_3$, respectively \cite{Iachello2006,Kota2020}. A mapping between the invariants of the two models \cite{Castanos1988,Draayer1989} leads to the connection between the collective variables $\beta$ and $\gamma$ and the Elliott quantum numbers $\lambda$ and $\mu$, given by \cite{Castanos1988,Draayer1989}
\begin{equation}\label{g1}
\gamma = \arctan \left( {\sqrt{3} (\mu+1) \over 2\lambda+\mu+3}  \right),
\end{equation}
and \cite{Castanos1988,Draayer1989}
\begin{equation}\label{b1}
	\beta^2= {4\pi \over 5} {1\over (A \bar{r^2})^2} (\lambda^2+\lambda \mu + \mu^2+ 3\lambda +3 \mu +3), 
\end{equation}
where $A$ is the mass number of the nucleus, while $\bar{r^2}$ is related to the dimensionless mean square radius \cite{Ring1980}, $\sqrt{\bar{r^2}}= r_0 A^{1/6}$. The dimensionless mean square radius is obtained by dividing the mean square radius, which grows as $A^{1/3}$, by the oscillator length, which is proportional to $A^{1/6}$ \cite{Ring1980}. The constant $r_0$ is found from a fit over a wide range of nuclei \cite{DeVries1987,Stone2014} to have the value $r_0=0.87$. The quantity appearing in the parentheses is related to the eigenvalues 
of the second order Casimir operator of SU(3), given by 
\cite{Kota2020}
  \begin{equation}\label{C2} 
 C_2(\lambda,\mu)= (\lambda^2+\lambda \mu + \mu^2+ 3\lambda +3 \mu). 
\end{equation}
We remark that $\beta^2$ is proportional to $C_2+3$. However, within the proxy-SU(3) framework, only the valence shells have been considered. Therefore, the values of $\beta$ should be multiplied by a scaling factor $A/(S_p+S_n)$, with $S_p$ ($S_n$) being the size of the proton (neutron) valence shell \cite{Bonatsos2017b}. For example, in the case of the actinides, in which the valence protons lie in the 82-126 shell and the valence neutrons lie in the 126-184 shell, one has $S_p=44$ and $S_n=58$, thus the scaling factor is $A/102$.

The numerical results for $\beta$ and $\gamma$ in the region with $Z=82$-126, $N=82$-258 are given in Appendix B. Although there are several theoretical estimates for the proton and neutron driplines in this region \cite{Wang2015,Neufcourt2020,Yang2021,Chai2022,Ahn2024}, the full numerical results have been reported for the sake of completeness, 
as well as for allowing the testing of mirror symmetries, as it will be seen in section V.B.
The following observations can be made. 

In the cases in which the hw irrep $(\lambda,\mu)$ is followed by the nhw irrep $(\lambda+2,\mu-4)$, the values of $\beta$ remain almost unchanged, since from Eq. (\ref{C2}) one can easily see that $C_{nhw}= C_{hw} -6 (\mu-1)$ and in addition in most cases the involved values of $\lambda$ are much larger than $\mu$. 

In the cases in which the hw irrep $(\lambda,\mu)$ is followed by the nhw irrep $(\lambda+2,\mu-4)$, the values of $\gamma$ are also moderately changed, since in Eq. (\ref{g1}) one can easily see that the denominator retains the same value in both the hw and the nhw irreps, while the numerator is changed 
from $(\mu+1)$ in the hw irrep to $(\mu-3)$ in the nhw irrep. Since the denominator contains $\lambda$, which is usually much larger than $\mu$, the change in $\gamma$ remains rather small. 

The above two points explain the robustness of the proxy-SU(3) predictions, as seen since its early applications \cite{Bonatsos2017b}. The hw irrep alone suffices to give reliable predictions for the collective variables $\beta$ and $\gamma$, while the involvement of the nhw irrep would have very minor influence.

The only exceptions to the above statement occur in the cases mentioned in Sec. \ref{irreps}, in which the valence proton or the valence neutron numbers acquire any of the values $M=2$, 4, 6, 12, 20, 30, 42, 56. In these cases the contribution of the nhw irrep can be sizeable, especially in the case of 
$\gamma$, as seen, for example, in nuclei in the rare earth region in Refs. \cite{Bonatsos2024b,Bonatsos2025}.

\section{Examples}

\subsection{Prolate to oblate shape transitions} \label{proobl}

In Fig. 1 the values of the $\beta$ and $\gamma$ variables are plotted for the nuclei with $Z=114$-124, i.e. the nuclei below $Z=126$, 
in the three different regions bordered by the neutron shell closures 82, 126, 184, 258.

For  $\beta$ a change in the slope of the curves is observed at  $N=112$, 168, and 240, with the $\beta$ values after these neutron numbers 
bending upwards, which could be a sign of a shape/phase transition \cite{Casten2006,Casten2007,Casten2009,Cejnar2009,Cejnar2010}.     

Indeed, we see that deep minima for $\gamma$ appear at the neutron numbers $N=112$, 168, and 240, while the values of $\gamma$ suddenly jump from prolate-like shapes 
(with $\gamma < 30^{\rm o}$) to oblate-like shapes (with $\gamma > 30^{\rm o}$). 

Such a transition has been seen in the rare earth region at $N=112$ below the $Z=82$ proton shell closure \cite{Bonatsos2017b}, as discussed in detail in the recent review article \cite{Bonatsos2024a}. 

It is worth remarking that this transition occurs at neutron numbers which are magic numbers of the 3D-HO. It is then of interest to examine if this ``remembrance'' of the 3D-HO magic numbers, seen in the proxy-SU(3) framework, is seen in other theoretical frameworks. Indeed, covariant density functional theory (CDFT) calculations involving the DD-ME2 functional in this region \cite{Bonatsos2022a,Bonatsos2022b}, show that the single particle energies of various orbitals converge at $N=112$, as they do at the shell closure at $N=126$. Examples can be seen for the $1h_{11/2}$, $1h_{9/2}$, $1g_{7/2}$  proton orbitals of Po, Pb, and Hg, in Figs. 1, 3, and 5 of Ref. \cite{Bonatsos2022b}.  Convergence in the $N=126$ case means degeneracy of the single particle energy levels because of the presence of spherical symmetry. It seems that the same spherical symmetry also appears at $N=112$. It should be emphasized that no magic numbers are built in  in mean-field theories in general, but they come out naturally from the interaction. In addition to providing microscopic support for the simultaneous presence of the usual (spin-orbit based) magic numbers and the 3D-HO magic numbers, these CDFT results also serve as confirmation of the dual shell mechanism \cite{Martinou2021,Martinou2023,Bonatsos2023a} for shape coexistence \cite{Heyde1983,Wood1992,Heyde2011,Heyde2016,Garrett2022,Bonatsos2023b} suggested within the proxy-SU(3) scheme. The CDFT predictions for the single-particle energy spectra have been recently corroborated by non-relativistic mean-field calculations using the Hartree-Fock-Bogoliubov theory \cite{Hasan2026}. 

It should be remarked that in Fig. 1 no nuclei with $\gamma=0$ appear. This is in agreement to recent microscopic Monte Carlo shell-model calculations showing \cite{Otsuka2025} that a small amount of triaxiality is seen even in well-deformed rare-earth nuclei considered as perfect prolate rotors up to now, in agreement with predictions made within the triaxial projected shell model \cite{Rouoof2024}, and with recent experimental evidence \cite{Kleemann2025} supporting this suggestion. Triaxiality within the proxy-SU(3) approach has been considered in detail in Refs. \cite{Bonatsos2024b,Bonatsos2025}, in which regions exhibiting high triaxiality have been singled out from the bulk of nuclei exhibiting low or moderate triaxiality. The role of the nhw irreps within the proxy-SU(3) framework has also been clarified \cite{Bonatsos2024b} through comparison to the Monte Carlo shell-model microscopic predictions.

Furthermore, the prolate over oblate dominance within the proxy-SU(3) framework has been discussed in the early papers \cite{Bonatsos2017a,Bonatsos2017b,Sarantopoulou2017}, as well as in Ref. \cite{Bonatsos2017c} within a more general framework. In these references, nuclei with $\gamma< 30^{\rm o}$ are considered as prolate-leaning, while nuclei with 
$\gamma > 30^{\rm o}$ are considered as oblate-leaning. The prolate over oblate dominance is based on the fact that the number of prolate-leaning nuclei is much larger than the number of oblate-leaning nuclei. 

\subsection{Mirror symmetry} \label{mirror}

In a recent study \cite{Zong2024}, the valence mirror symmetry between the $Z=50$ isotopes and the $N=82$ isotones has been considered in the framework of the nucleon-pair approximation (NPA) \cite{Zhao2014}. We are going to examine to what extent such a mirror symmetry occurs just above the $Z=82$ and $N=126$ magic numbers. 

In Fig. 2(a) the proxy-SU(3) predictions for the collective variable $\beta$ are shown, considering up to three valence proton pairs above the $Z=82$ shell closure, i.e., $Z=84$, 86, 88 and $N=84$-124, as well as up to three valence neutron pairs above the $N=126$ shell closure, i.e.  $N=128$, 130, 132 and $Z=84$-124.   Very close agreement is seen between the $Z=84$ and $N=130$ predictions, which regard nuclei with two protons outside the $Z=82$ shell and nuclei with four neutrons outside the $N=126$ shell. Furthermore, very close agreement is seen between the $Z=86$ and $N=132$ predictions, i.e., between nuclei with four protons outside the $Z=82$ shell and nuclei with six neutrons outside the $N=126$ shell.
The agreement is better especially around the middle of the 82-126 interval, where maximum quadrupole deformation occurs.

It is of interest to examine if this similarity also appears for the corresponding holes. The relevant plot is seen in Fig. 2(b).
Indeed we see that similarity is seen between the $Z=124$ and $N=180$ predictions, i.e., for nuclei with two protons holes below the $Z=126$ shell and nuclei with four neutrons holes below the $N=184$ shell,  as well as between the $Z=122$ and $N=178$ predictions, i.e., for nuclei with four protons holes below the $Z=126$ shell and nuclei with six neutrons holes below the $N=184$ shell. The similarities are seen especially below ${\bf M}=112$ particles. However, the $\beta$ values occurring in the holes cases in Fig. 2(b) are lower than the ones occurring in the corresponding particles cases in Fig. 2(a). 
 
In conclusion, we see that a certain degree of similarity exists between the nuclei with two (four) protons outside the proton closed  shell and the nuclei with four (six) neutrons outside the neutron closed shell, as well as between the nuclei with two (four) proton holes inside the proton closed closed  shell and the nuclei with four (six) neutron holes inside the neutron closed shell. However, in the case of holes the predicted values of $\beta$ are systematically lower than the corresponding values for particles. In addition, the similarities are stronger below the 3D-HO magic number 112 than above it. The microscopic roots of these two observations are calling for further investigation.  

It would have been very interesting to examine if this mirror symmetry appears also in mean-field calculations. However, in most cases the published results 
are limited within the drip lines, thus preventing such a test. For example, in the extended tables of mean-field predictions obtained with the D1S Gogny interaction (see Ref. [1] of \cite{Delaroche2010}), and in relation to Fig. 2(a) of our manuscript, the Gogny D1S tables provide the following sets of results: a) For N=128,130,132 they provide results for Z=84-96, corresponding to the left part of Fig. 2(a), b) For Z=84,86,88 they provide results for N=108-124, corresponding to the right part of Fig. 2(a). There is no overlap between these two regions, thus the analog of Fig. 2(a) cannot be drawn based on the existing results, just because they are limited to the region between the driplines. The same holds for Fig. 2(b). However, similar tests of the mirror symmetry within the D1S Gogny approach are possible in the rare-earth region, in which sufficient numerical results exist in the region between the drip lines and should be pursued. 

\subsection{Evolution of collective variables along the valley of stability} \label{valley} 

A test for the proxy-SU(3) predictions can be provided by the valley of stability, recently used as a test-ground for a potential-energy-surface (PES) approach \cite{Meng2022}. 

The valley of stability is given by Green's formula \cite{Green1955}
\begin{equation}
N-Z = 0.4 {A^2 \over A+200}.  
\end{equation}  
The nuclei along the stability line are listed in Table II. 

The proxy-SU(3) predictions for the collective variable $\beta$ along the valley of stability, obtained with the hw irrep, are shown in Fig. 3(a). The empirical values shown are obtained from experimental data for the transition rates $B(E2; 0_1^+\to 2_1^+)$ \cite{Pritychenko2016}. Good agreement is seen between the proxy-SU(3) predictions and the empirical values, especially in the region above $Z=82$. 

The proxy-SU(3) predictions for the collective variable $\gamma$ along the valley of stability, obtained with the hw irrep, are shown in Fig. 3(b).
 We see that deep minima occur at the proton numbers 32, 40, 52, 62, 70, 84, 94, 112. These are related to the collection of 2, 4, 6, 12, 20, 30 valence protons, related to irreps with $\mu=0$, as seen in Sec. \ref{irreps}. The first two numbers, 32 and 40, represent 4 and 12 protons above the magic number 28, while the next three numbers, 52, 62, and 70, represent    2, 12, and 20 protons above the magic number 50.  Finally, the last three numbers, 84, 94, and 112, represent  2, 12, 30 protons above the magic number 82.
 
 As discussed in Sec. \ref{coll} and Refs. \cite{Bonatsos2024b,Bonatsos2025}, the proxy-SU(3) predictions for the collective variable $\gamma$ at the deep minima, should be replaced by the average value of $\gamma$ obtained from the hw irrep and the nhw irrep, resulting in a smoother curve, as seen in Fig. 3(c). 
Relevant predictions by the potential energy surface (PES) approach are given in the middle panel of Fig. 3 of Ref. \cite{Meng2022}. Comparison between the two figures reveals that the maxima of gamma occur at nearly the same $Z$ numbers, namely at 36, 54, 80, 110 in the present work and at 34, 52, 80, 110 in Ref. \cite{Meng2022}. Since the results come from completely different theoretical approaches, this is an encouraging result.

\section{Summary and outlook} 

The hw irreps and nhw SU(3) irreps in the framework of the proxy-SU(3) approximation to the shell model have been determined for all nuclei in the region 
of $Z$=82-126, $N$=82-258, and the corresponding parameter-free predictions for the $\beta$ and $\gamma$ collective variables are given. The numerical results have been used for studying the prolate to oblate shape transition, for testing mirror symmetries in heavy nuclei, as well as for examining the evolution of the collective variables along the valley of stability.

Since numerical results for the collective variables $\beta$ and $\gamma$ from several mean-field approaches exist, along with some experimental data, especially in the actinide region, a systematic comparison between them and the present proxy-SU(3) predictions should be carried out. Work in this direction is in progress.  

The numerical results for the collective variables $\beta$ and $\gamma$ presented in this article regard the actinides ($82 < Z < 104$) and the superheavy nuclei ($104\leq Z \leq 126$). Results for hyperheavy nuclei ($Z>126$) can be obtained  in a straightforward way from  Eqs. (\ref{g1})
and (\ref{b1}), and the irreps listed in Table I, which are extended up to 350 protons or neutrons. However, one should bear in mind that beyond 
$Z=126$ and $N=258$ totally different shapes might appear, including toroidal ones, as predicted by relativistic mean-field calculations 
\cite{Agbemava2019, Afanasjev2020,Agbemava2021}.    

A point which seems to justify further attention is the appearance of magic numbers of the three-dimensional isotropic harmonic oscillator (3D-HO) as playing a role in the transition from prolate to oblate shapes, as well as in the dual shell mechanism for shape coexistence, discussed in Sec. V.A. While the proxy-SU(3) shells result by taking into account the spin-orbit interaction \cite{Martinou2020}, and thus they are separated by the standard shell-model magic numbers, the parameter-independent numerical results for the collective variables obtained within the proxy-SU(3) framework exhibit some ``remembrance'' of the 3D-HO magic numbers, playing a decisive role on the neutron number at which the prolate to oblate transition is observed within a series of isotopes, as well as on the upper border of neutron and proton regions in which the appearance of shape coexistence is predicted to be possible by the dual shell mechanism \cite{Martinou2021} suggested within the proxy-SU(3) framework. What is even more interesting, is that convergence of single-particle energy levels into a single degenerate level is observed at the 3D-HO magic numbers within mean-field calculations, both in the relativistic \cite{Bonatsos2022a,Bonatsos2022b} and non-relativistic \cite{Hasan2026} frameworks. Since in mean-field theories no magic numbers are built in by construction, the appearance of these degeneracies, which are hallmarks of symmetries, is a rather surprising feature, the consequences of which call for further investigation. 

\section*{Acknowledgements} 

This research project is implemented in the framework of the Hellenic Foundation for Research and Innovation (H.F.R.I.) call ``3rd Call for H.F.R.I.'s Research Projects to Support Faculty Members and Researchers'' (H.F.R.I. Project Number: 23357). Support by the Bulgarian National Science Fund (BNSF) under Contract No. KP-06-N98/2 is gratefully acknowledged.

\section*{Appendix A}

In this Appendix some regularities appearing in the entries of Table I are described. These regularities connect the hw and nhw irreps 
of a unitary algebra appearing in Table I to the hw and nhw irreps of the next unitary algebra appearing in Table I.

In Table I the hw SU(3) irreps and the nhw SU(3) irreps are given for the unitary algebras U$((\eta+1)(\eta+2)/2)$ with $\eta=2$, 3, 4, 5, 6, 7, 8. 

We remark that if for a given number of particles $M$ and for $\eta=n$ the hw irrep  is $(\lambda_{hw}, \mu_{hw})$, then for $\eta=n+1$
the hw irrep is  $(\lambda_{hw}+M, \mu_{hw})$.

We also remark that if for a given number of particles $M$ and for $\eta=n$ the nhw irrep  is $(\lambda_{nhw}, \mu_{nhw})$, then for $\eta=n+1$
the nhw irrep is  $(\lambda_{nhw}+M, \mu_{nhw})$.

These regularities can be used for the extension of Table I to higher shells, without need of using the code UNTOU3 \cite{Draayer1989a}.

\section*{Appendix B}

In this Appendix the highest weight (hw) irreducible representations (irreps) of SU(3) and the next highest weight (nhw) irreps of SU(3) within the proxy-SU(3) scheme for nuclei in the $Z=84$-124, $N=84$-256 region are tabulated, in order of increasing $Z$, labeled by the Elliott \cite{Elliott1958a} notation $(\lambda,\mu)$. The parameter-independent proxy-SU(3) 
predictions for the collective variables $\beta$ and $\gamma$, calculated from Eqs. (\ref{b1}) and (\ref{g1}), are also given for each nucleus, labeled by hw and nhw respectively.

\include{ex1481}

%%%%%%%%%%%%%%%%%%%%%%%%%%%%%%%%%%%%%%%%%%%%%%%%%%%%%%%%%%%%%%%%%%%%%%%%%%%%%%%%%%%%%%%%%%%%%%%%%%%%%%%%%%%%%%%%%%%%%%%%%%%%%%%%%%%%%%%%%%%%%%%%   

%%% Table 1 %%%%%%%%%%%%%%
\setcounter{table}{0}
\begin{table*}

\caption{Highest weight (hw) irreducible representations of SU(3) and next highest weight (nhw) irreps of SU(3) for $M$ nucleons within the proxy-SU(3) scheme in the $sd$, $pf$, $sdg$, $pfh$, $sdgi$, $pfhj$, $sdgik$ shells having the overall symmetry  U(6), U(10), U(15), U(21), U(28),  U(36) and U(45) respectively, calculated using the code UNTOU3 \cite{Draayer1989a}. The Elliott \cite{Elliott1958a} notation $(\lambda,\mu)$ is used for the SU(3) irreps. The corresponding shells of the shell model, within the proxy-SU(3) scheme \cite{Martinou2020} are also shown. Partly adapted from Ref. \cite{Bonatsos2024b}. See Section \ref{irreps} and Appendix A for further discussion. 
}
\begin{tabular}{ r r r r r r r r r r r r r r r  }
\hline
$M$  & U(6) & U(6) & U(10) & U(10) & U(15) & U(15) & U(21) & U(21) & U(28) & U(28) & U(36) & U(36) & U(45) & U(45) \\
     & $sd$ & $sd$ & $pf$  & $pf$  & $sdg$ & $sdg$ & $pfh$ & $pfh$ & $sdgi$ & $sdgi$ & $pfhj$ & $pfhj$ & $sdgik$ & $sdgik$ \\
     & 8-20 & 8-20 & 28-50 & 28-50 & 50-82 & 50-82 & 82-126 & 82-126 & 126-184 & 126-184 & 184-258 & 184-258 & 258-350  & 258-350  \\
     & hw   & nhw  & hw    & nhw   & hw    & nhw   & hw    & nhw   & hw    & nhw  & hw  &  nhw  & hw & nhw \\

      \hline
 2 & 4,0 & 0,2 &  6,0 &  2,2 &  8,0 &   4,2 &  10,0 &   6,2 &  12,0 &   8,2 &  14,0 &  10,2 &   16,0 &   12,2 \\ 
 4 & 4,2 & 0,4 &  8,2 &  4,4 & 12,2 &   8,4 &  16,2 &  12,4 &  20,2 &  16,4 &  24,2 &  20,4 &   28,2 &   24,4 \\
 6 & 6,0 & 0,6 & 12,0 &  6,6 & 18,0 &  12,6 &  24,0 &  18,6 &  30,0 &  24,6 &  36,0 &  30,6 &   42,0 &   36,6 \\
 8 & 2,4 & 4,0 & 10,4 & 12,0 & 18,4 &  20,0 &  26,4 &  28,0 &  34,4 &  36,0 &  42,4 &  44,0 &   50,4 &   52,0 \\
10 & 0,4 & 2,0 & 10,4 & 12,0 & 20,4 &  22,0 &  30,4 &  32,0 &  40,4 &  42,0 &  50,4 &  52,0 &   60,4 &   62,0 \\
12 & 0,0 &     & 12,0 & 4,10 & 24,0 & 16,10 &  36,0 & 28,10 &  48,0 & 40,10 &  60,0 & 52,10 &   72,0 &  64,10 \\
14 &     &     &  6,6 &  8,2 & 20,6 &  22,2 &  34,6 &  36,2 &  48,6 &  50,2 &  62,6 &  64,2 &   76,6 &   78,2 \\
16 &     &     &  2,8 &  4,4 & 18,8 &  20,4 &  34,8 &  36,4 &  50,8 &  52,4 &  66,8 &  68,4 &   82,8 &   84,4 \\
18 &     &     &  0,6 &  2,2 & 18,6 &  20,2 &  36,6 &  38,2 &  54,6 &  56,2 &  72,6 &  74,2 &   90,6 &   92,2 \\
20 &     &     &  0,0 &      & 20,0 & 10,14 &  40,0 & 30,14 &  60,0 & 50,14 &  80,0 & 70,14 &  100,0 &  90,14 \\
22 &     &     &      &      & 12,8 &  14,4 &  34,8 &  36,4 &  56,8 &  58,4 &  78,8 &  80,4 &  100,8 &  102,4 \\
24 &     &     &      &      & 6,12 &   8,8 & 30,12 &  32,8 & 54,12 &  56,8 & 78,12 &  80,8 & 102,12 &  104,8 \\
26 &     &     &      &      & 2,12 &   4,8 & 28,12 &  30,8 & 54,12 &  56,8 & 80,12 &  82,8 & 106,12 &  108,8 \\
28 &     &     &      &      &  0,8 &   2,4 &  28,8 &  30,4 &  56,8 &  58,4 &  84,8 &  86,4 &  112,8 &  114,4 \\
30 &     &     &      &      &  0,0 &       &  30,0 & 18,18 &  60,0 & 48,18 &  90,0 & 78,18 &  120,0 & 108,18 \\
32 &     &     &      &      &      &       & 20,10 &  22,6 & 52,10 &  54,6 & 84,10 &  86,6 & 116,10 &  118,6 \\
34 &     &     &      &      &      &       & 12,16 & 14,12 & 46,16 & 48,12 & 80,16 & 82,12 & 114,16 & 116,12 \\ 
36 &     &     &      &      &      &       &  6,18 &  8,14 & 42,18 & 44,14 & 78,18 & 80,14 & 114,18 & 116,14 \\
38 &     &     &      &      &      &       &  2,16 &  4,12 & 40,16 & 42,12 & 78,16 & 80,12 & 116,16 & 118,12 \\
40 &     &     &      &      &      &       &  0,10 &   2,6 & 40,10 &  42,6 & 80,10 &  82,6 & 120,10 &  122,6 \\
42 &     &     &      &      &      &       &   0,0 &       &  42,0 & 28,22 &  84,0 & 70,22 &  126,0 & 112,22 \\
44 &     &     &      &      &      &       &       &       & 30,12 &  32,8 & 74,12 &  76,8 & 118,12 &  120,8 \\
46 &     &     &      &      &      &       &       &       & 20,20 & 22,16 & 66,20 & 68,16 & 112,20 & 114,16 \\
48 &     &     &      &      &      &       &       &       & 12,24 & 14,20 & 60,24 & 62,20 & 108,24 & 110,20 \\
50 &     &     &      &      &      &       &       &       &  6,24 &  8,20 & 56,24 & 58,20 & 106,24 & 108,20 \\
52 &     &     &      &      &      &       &       &       &  2,20 &  4,16 & 54,20 & 56,16 & 106,20 & 108,16 \\
54 &     &     &      &      &      &       &       &       &  0,12 &   2,8 & 54,12 &  56,8 & 108,12 &  110,8 \\
56 &     &     &      &      &      &       &       &       &   0,0 &       &  56,0 & 40,26 &  112,0 &  96,26 \\
58 &     &     &      &      &      &       &       &       &       &       & 42,14 & 44,10 & 100,14 & 102,10 \\
60 &     &     &      &      &      &       &       &       &       &       & 30,24 & 32,20 &  90,24 &  92,20 \\
62 &     &     &      &      &      &       &       &       &       &       & 20,30 & 22,26 &  82,30 &  84,26 \\
64 &     &     &      &      &      &       &       &       &       &       & 12,32 & 14,28 &  76,32 &  78,28 \\
66 &     &     &      &      &      &       &       &       &       &       &  6,30 &  8,26 &  72,30 &  74,26 \\
68 &     &     &      &      &      &       &       &       &       &       &  2,24 &  4,20 &  70,24 &  72,20 \\ 
70 &     &     &      &      &      &       &       &       &       &       &  0,14 &  2,10 &  70,14 &  72,10 \\
72 &     &     &      &      &      &       &       &       &       &       &   0,0 &       &   72,0 &  54,30 \\
74 &     &     &      &      &      &       &       &       &       &       &       &       &  56,16 &  58,12 \\
76 &     &     &      &      &      &       &       &       &       &       &       &       &  42,28 &  44,24 \\ 
78 &     &     &      &      &      &       &       &       &       &       &       &       &  30,36 &  32,32 \\
80 &     &     &      &      &      &       &       &       &       &       &       &       &  20,40 &  22,36 \\
82 &     &     &      &      &      &       &       &       &       &       &       &       &  12,40 &  14,36 \\
84 &     &     &      &      &      &       &       &       &       &       &       &       &   6,36 &   8,32 \\
86 &     &     &      &      &      &       &       &       &       &       &       &       &   2,28 &   4,24 \\
88 &     &     &      &      &      &       &       &       &       &       &       &       &   0,16 &   2,12 \\
90 &     &     &      &      &      &       &       &       &       &       &       &       &    0,0 &        \\

\hline
\end{tabular}

\end{table*}

%%% Table 2 %%%%%%%%%%%%%%%%%%%%%%%

\begin{table}

\caption{
The valley of stability, obtained from Green's formula \cite{Green1955}. See Sec. \ref{valley} for further discussion.  
}
\begin{tabular}{ r r r r r r r r r r r r r r r r    }
\hline
    &    &    &    &    &    &     &     &     &     &     &     &     &     &     &     \\
$Z$ & 30 & 32 & 34 & 36 & 38 & 40  & 42  & 44  & 46  & 48  & 50  & 52  & 54  & 56  & 58  \\
$N$ & 38 & 40 & 42 & 46 & 48 & 52  & 54  & 58  & 62  & 64  & 68  & 70  & 74  & 78  & 80  \\
    &    &    &    &    &    &     &     &     &     &     &     &     &     &     &     \\
$Z$ & 60 & 62 & 64 & 66 & 68 &  70 &  72 &  74 &  76 &  78 &  80 &  82 & 84  & 86  & 88  \\
$N$ & 84 & 88 & 92 & 84 & 98 & 102 & 106 & 110 & 114 & 118 & 120 & 124 & 128 & 132 & 136 \\
    &    &    &    &    &    &     &     &     &     &     &     &     &     &     &     \\
$Z$ &  90 &  92 &  94 &  96 &  98 & 100 & 102 & 104 & 106 & 108 & 110 & 112 & 114 & 116 & 118 \\
$N$ & 140 & 142 & 146 & 150 & 154 & 158 & 162 & 166 & 170 & 174 & 178 & 182 & 186 & 190 & 194 \\  
    &    &    &    &    &    &     &     &     &     &     &     &     &     &     &     \\

\hline
\end{tabular}

\end{table}

%%%%%%%%%%%%%%%%%%%%%%%%%%%%%%%%%%%%%%%%%%% FIG. 1  %%%%%%%%%%%%%%%%%%%%%%%%%%%%%%%%%%%%%%%%%%%%%

\begin{figure*} [htb]

{\includegraphics[width=75mm]{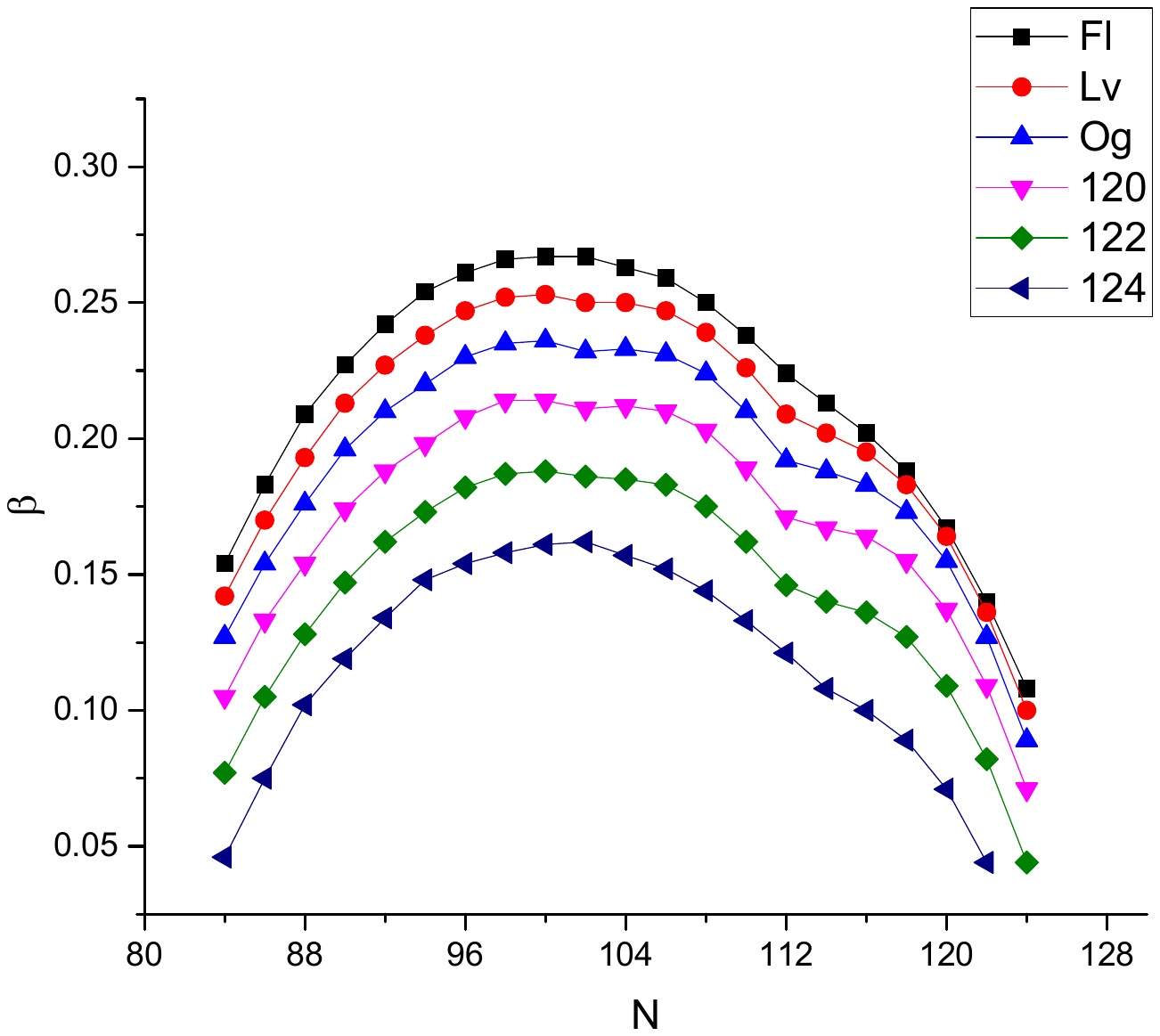}  \hspace{5mm}
    \includegraphics[width=75mm]{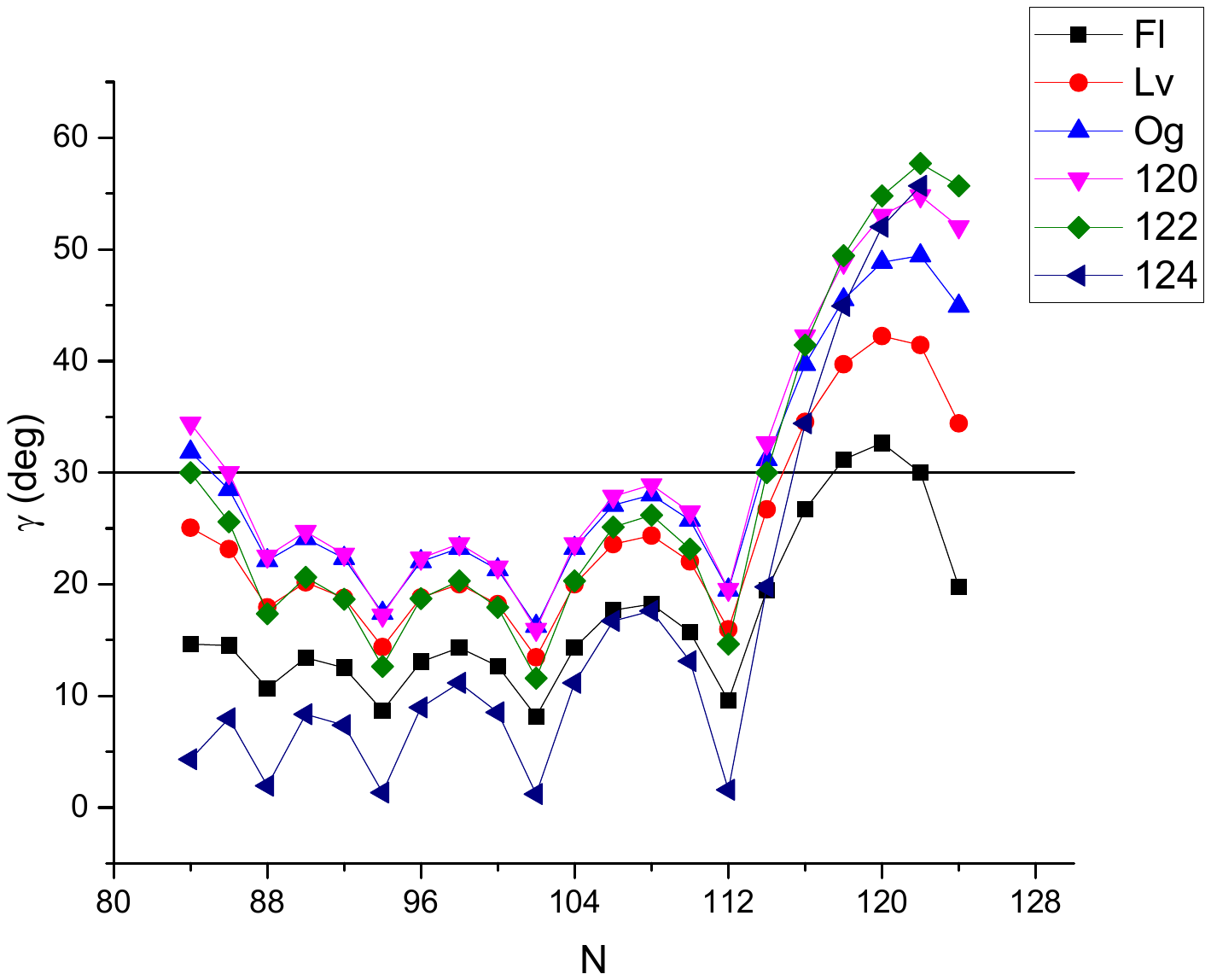}}
    {\includegraphics[width=75mm]{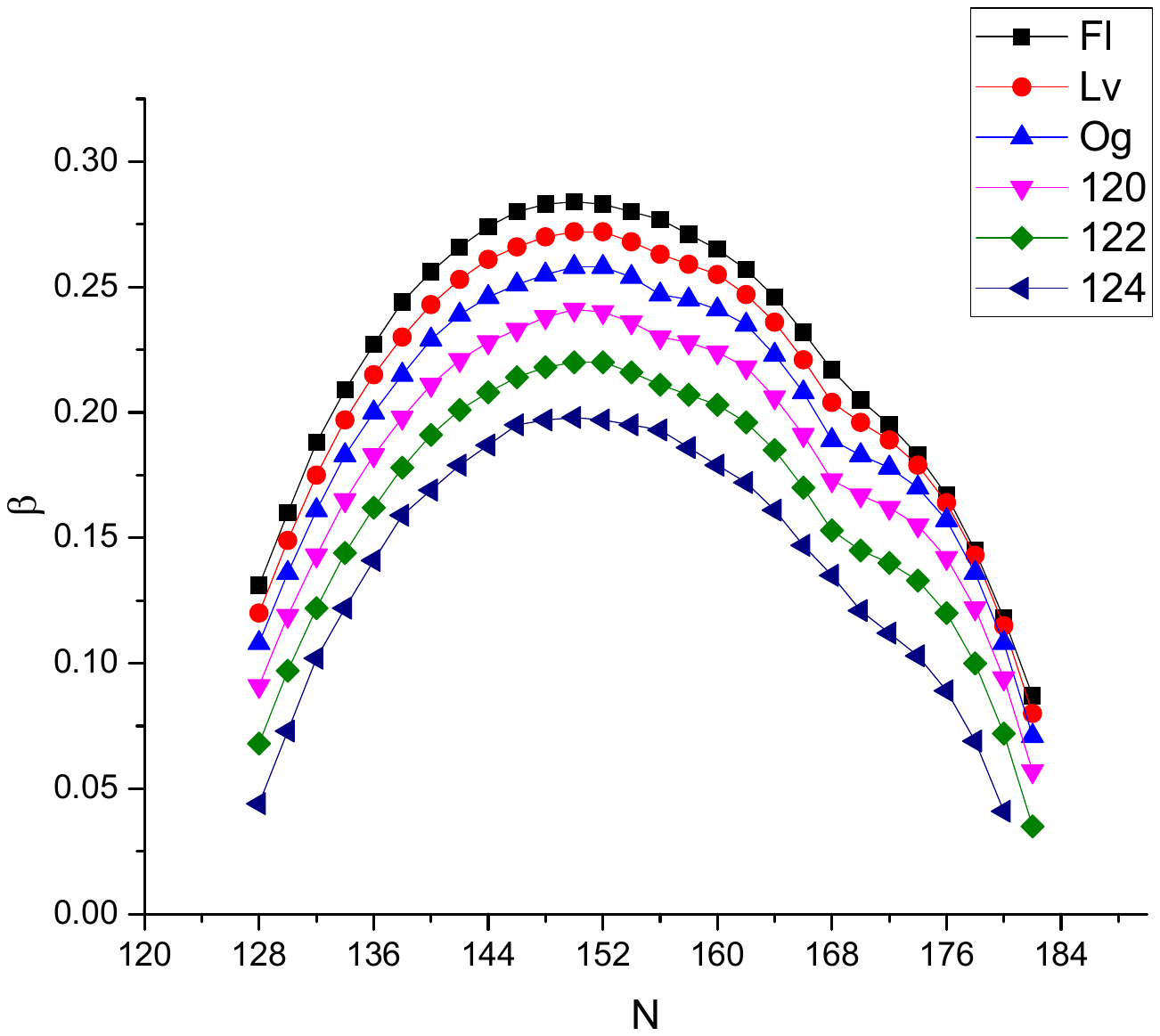}  \hspace{5mm}
    \includegraphics[width=75mm]{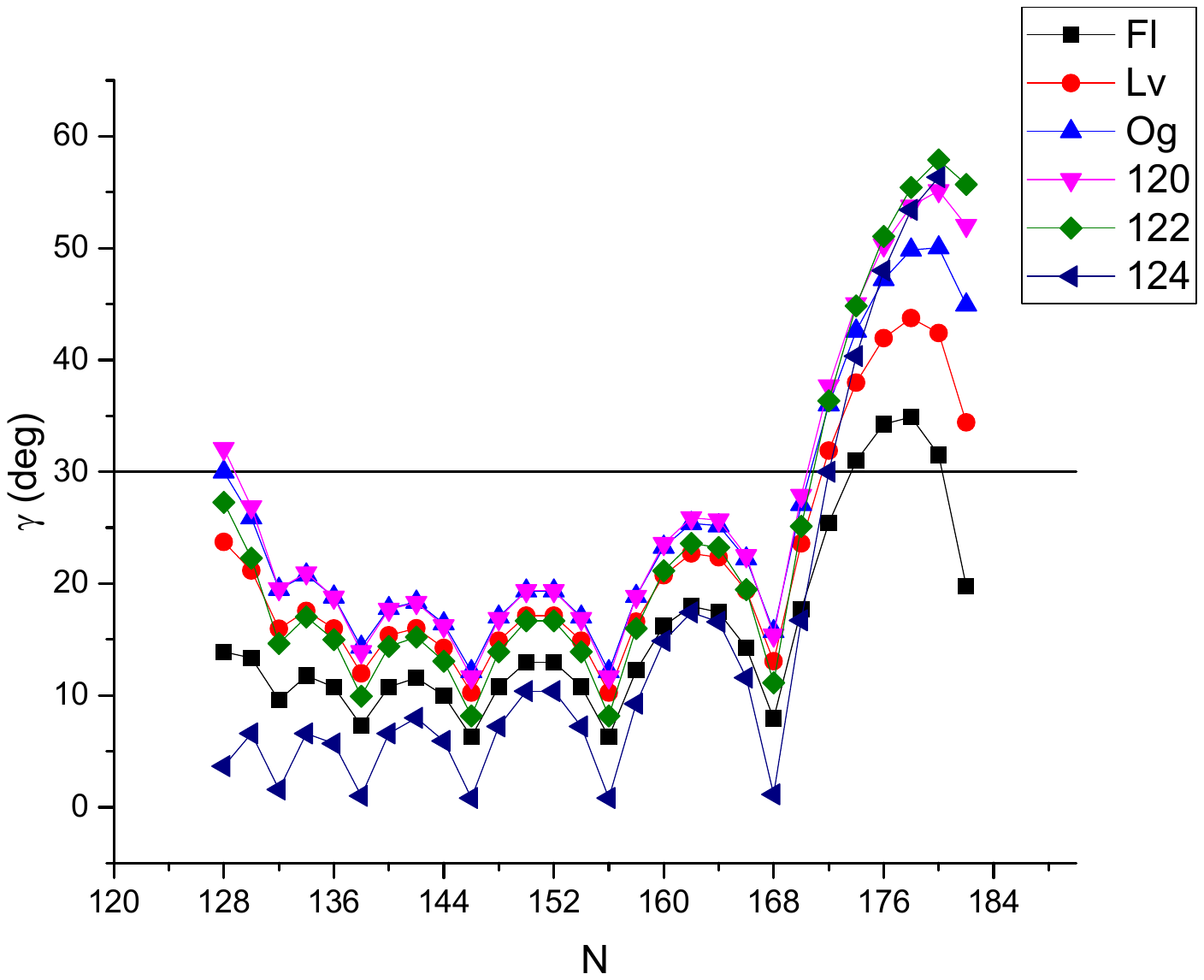}}
    {\includegraphics[width=75mm]{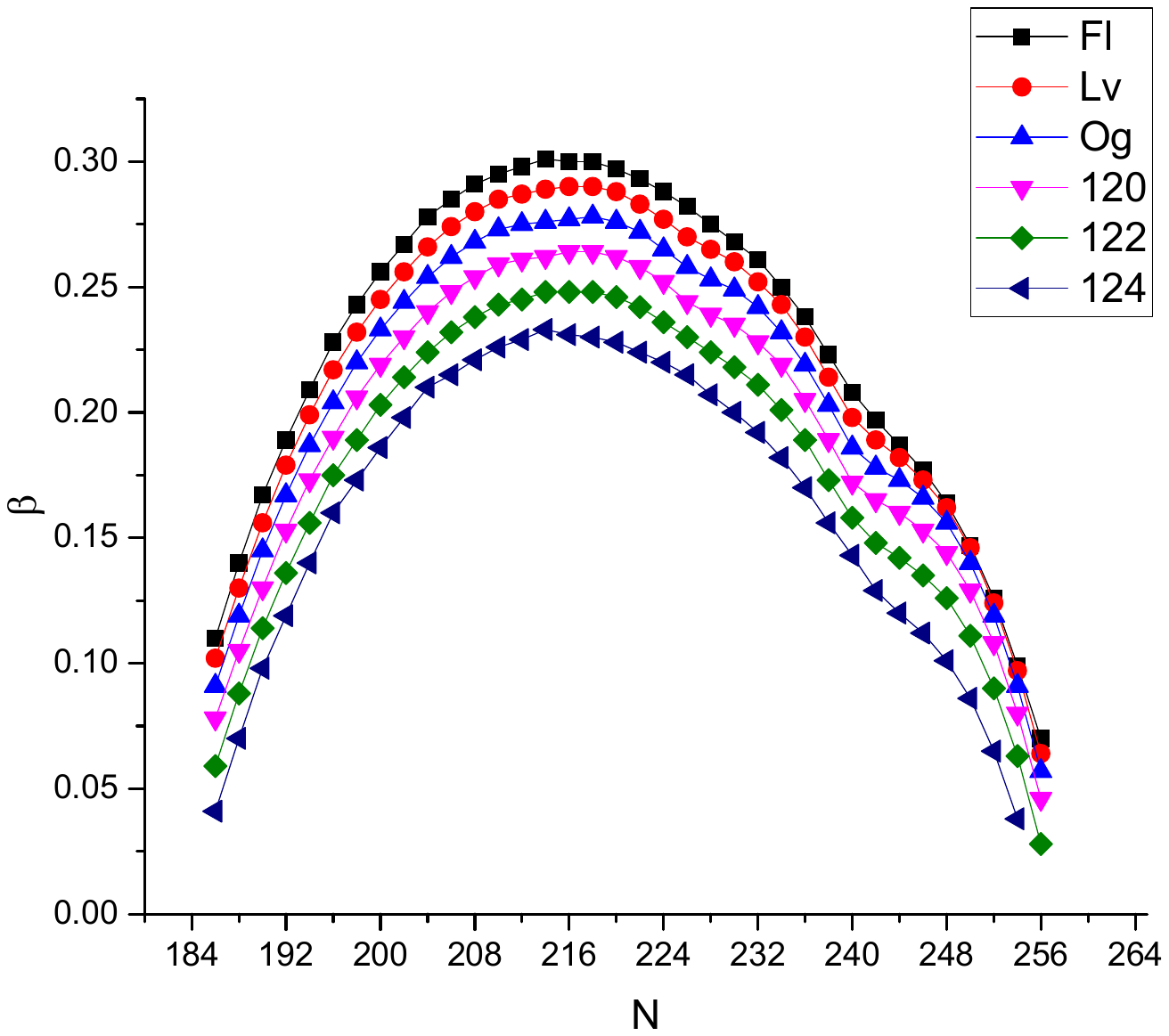}  \hspace{5mm}
    \includegraphics[width=75mm]{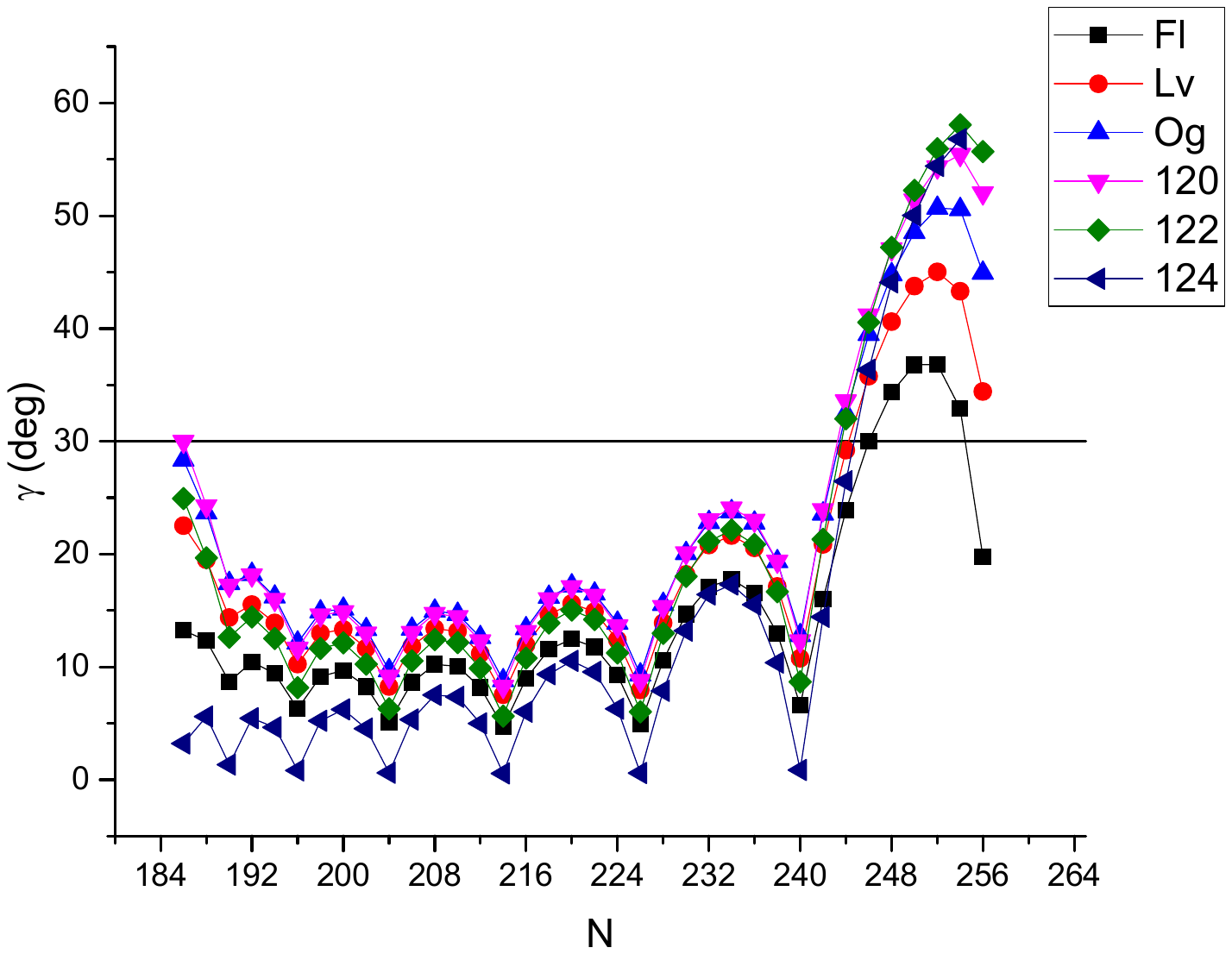}}
     
  \caption{Proxy-SU(3) predictions for the collective variables $\beta$ and $\gamma$ in the $Z=114$-124 region. See Sec. \ref{proobl} for further discussion. }
   
\end{figure*}

%%%%%%%%%%%%%%%%%%%%%%%%%%%%%%%%%%%%%%%%%%% FIG. 2 %%%%%%%%%%%%%%%%%%%%%%%%%%%%%%%%%%%%%%%%%%%%%

\begin{figure} [htb]

    \includegraphics[width=75mm]{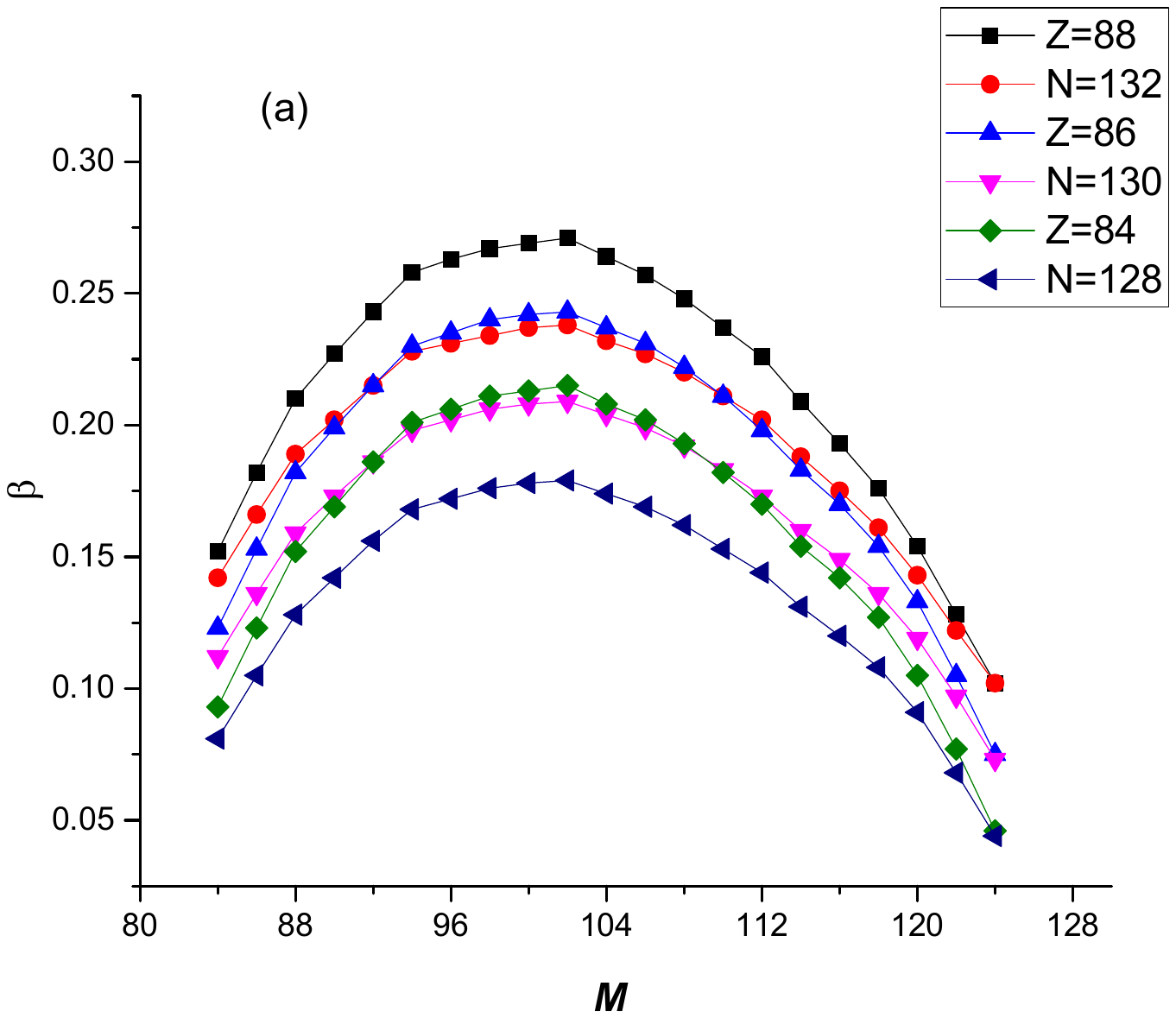} 
    \includegraphics[width=75mm]{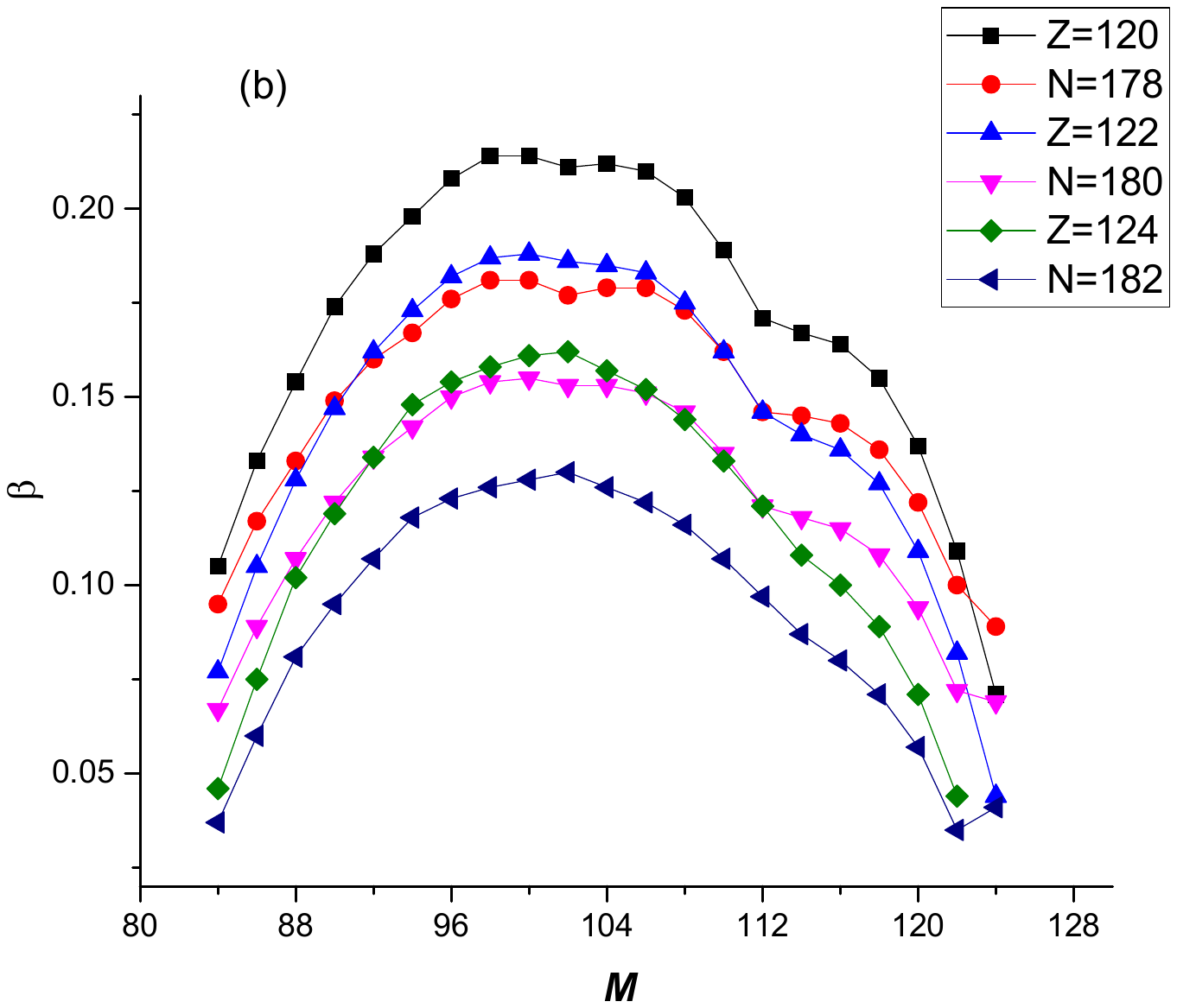} 
     
    \caption{(a) The parameter-free predictions for the collective variable $\beta$ obtained with the hw irrep of proxy-SU(3) for $Z=84$, 86, 88 (in which case 
    ${\bf M}$ stands for the neutron number $N$) and $N=128$, 130, 132 (in which case ${\bf M}$ stands for the proton number $Z$). 
b) Same for $Z=124$, 122, 120 and $N=182$, 180, 178. See Sec. \ref{mirror} for further discussion. }
 
\end{figure}

%%%%%%%%%%%%%%%%%%%%%%%%%%%%%%%%%%%%%%%%%%% FIG. 3 %%%%%%%%%%%%%%%%%%%%%%%%%%%%%%%%%%%%%%%%%%%%%

\begin{figure} [htb]

    \includegraphics[width=75mm]{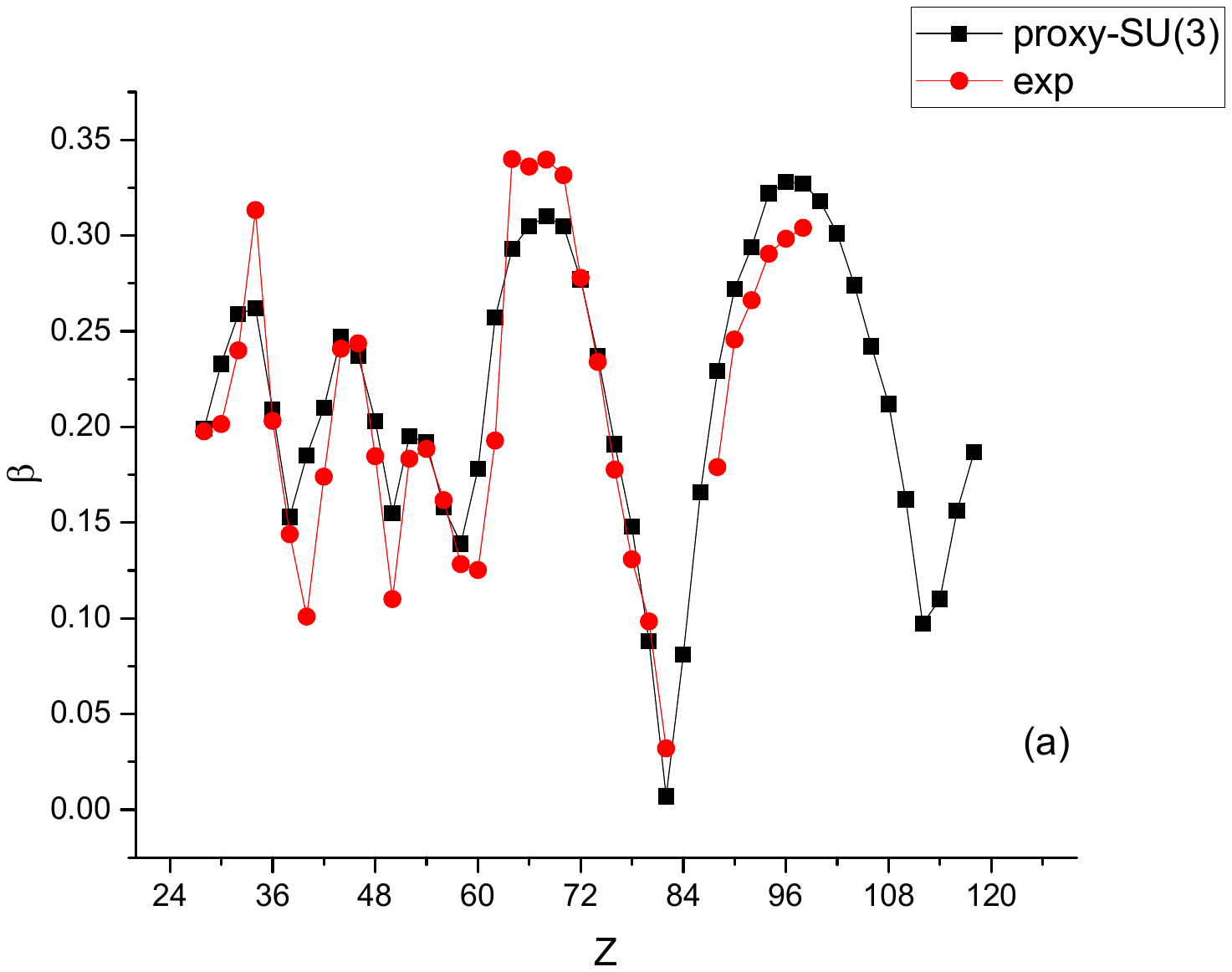} 
    \includegraphics[width=75mm]{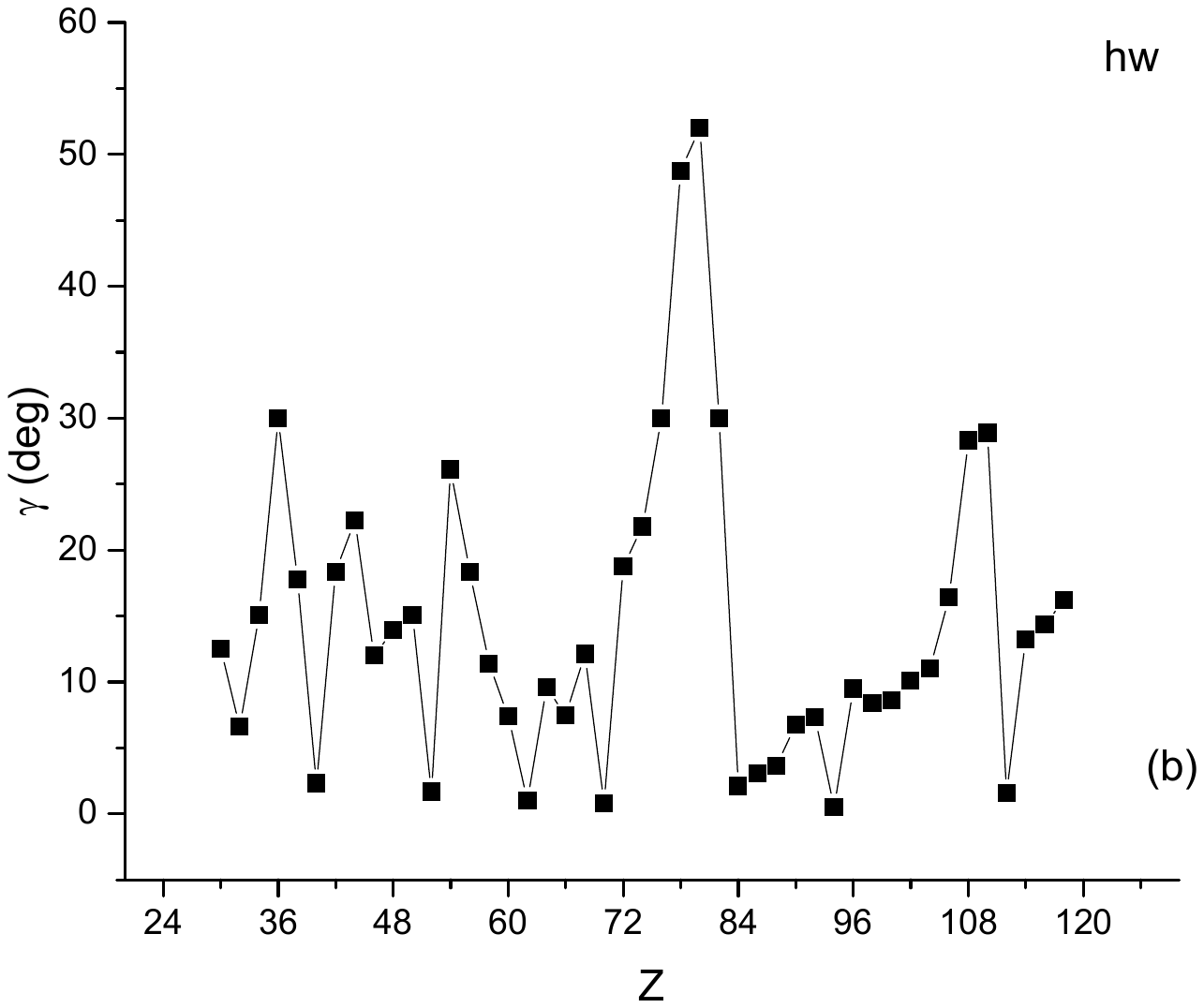} 
    \includegraphics[width=75mm]{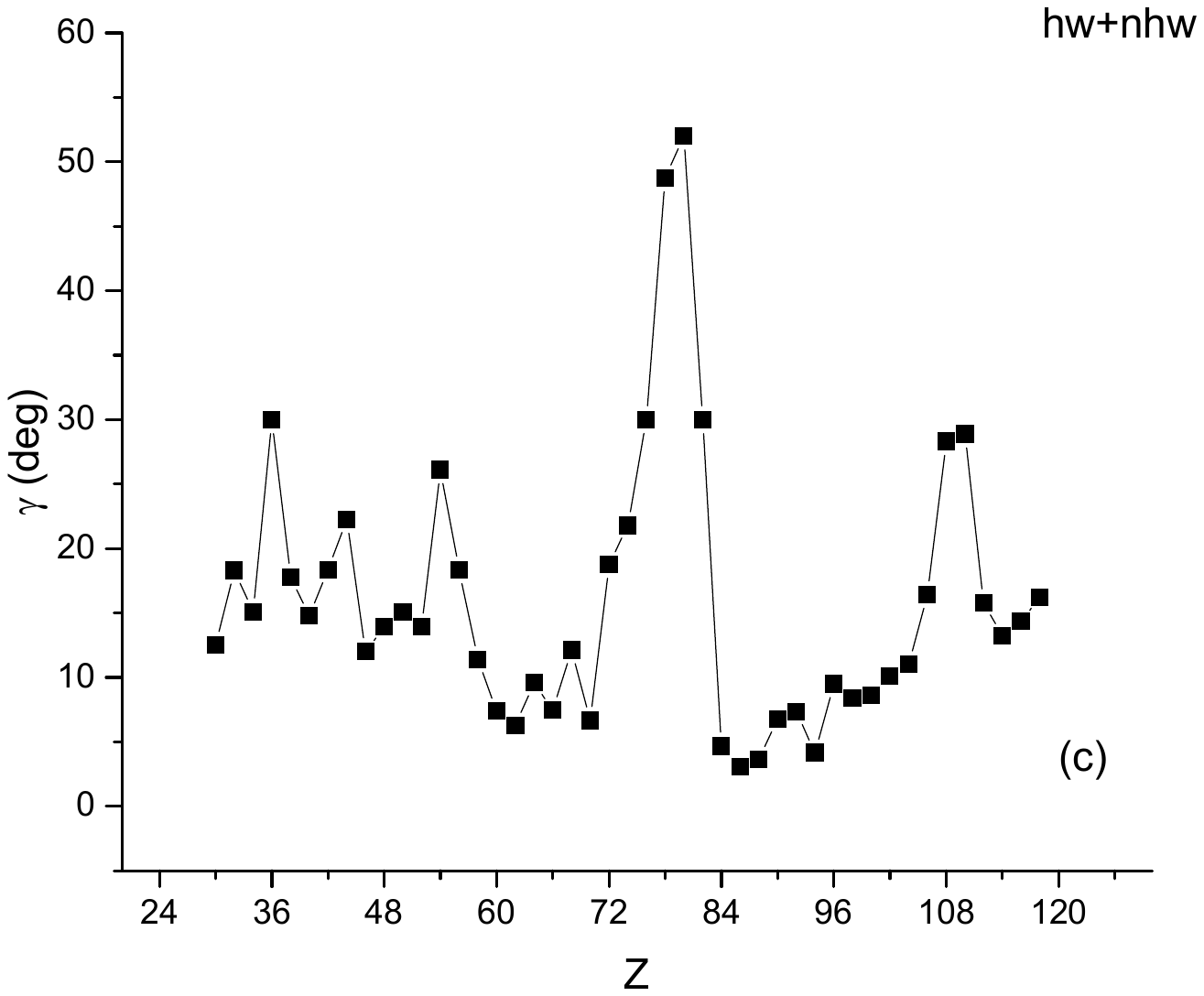} 
     
    \caption{(a) The parameter-free predictions for the collective variable $\beta$ along the valley of stability of Table II \cite{Meng2022,Green1955}, obtained with the hw irrep of proxy-SU(3), are compared to the empirical values taken from Ref. \cite{Pritychenko2016}. (b) Parameter-free predictions for the collective variable $\gamma$ along the valley of stability, obtained with the hw irrep of proxy-SU(3).
(c) Parameter-free predictions for the collective variable $\gamma$ along the valley of stability, obtained after mixing the hw and nhw irreps of proxy-SU(3). See Sec. \ref{valley} for further discussion.}

\end{figure}

\end{document}

%% file: ex1481.tex
%%%%%%%%%%%%%%%%%%%%%%%%%%%%%%%%%%%%%%%%%%%%%%%%%%%%%%%%%%%%%%%%%%%%%%%%%%%%%%%%%%%%%%%%%%%%%%%%%%%%%%%%%%%%%%%%%%%%%%%%%%%%%%%%%%%%%%%%%%%%%%%%   

\begin{table}

\begin{tabular}{ r   r  r r c c r  }
\hline
nucleus  & hw & nhw & $\beta_{hw}$ & $\beta_{nhw}$ & $\gamma_{hw}$ & $\gamma_{nhw}$  \\ 
        
\hline

 \isotope[168][84]{Po}$_{84}$ & 20,  0 & 16,  2 & 0.093 & 0.081 &  2.31 &  7.99 \\
 \isotope[170][84]{Po}$_{86}$ & 26,  2 & 22,  4 & 0.123 & 0.111 &  5.21 &  9.64 \\
 \isotope[172][84]{Po}$_{88}$ & 34,  0 & 28,  6 & 0.152 & 0.142 &  1.40 & 10.57 \\
 \isotope[174][84]{Po}$_{90}$ & 36,  4 & 38,  0 & 0.169 & 0.168 &  6.26 &  1.26 \\
 \isotope[176][84]{Po}$_{92}$ & 40,  4 & 42,  0 & 0.186 & 0.185 &  5.69 &  1.14 \\
 \isotope[178][84]{Po}$_{94}$ & 46,  0 & 38, 10 & 0.201 & 0.193 &  1.05 & 12.08 \\
 \isotope[180][84]{Po}$_{96}$ & 44,  6 & 46,  2 & 0.206 & 0.205 &  7.13 &  3.07 \\
\hline
\end{tabular}

\end{table}

\setcounter{table}{1} 
\begin{table}

\begin{tabular}{ r   r  r r c c r  }
\hline
nucleus  & hw & nhw & $\beta_{hw}$ & $\beta_{nhw}$ & $\gamma_{hw}$ & $\gamma_{nhw}$  \\ 
        
\hline

\isotope[270][84]{Po}$_{186}$ & 24,  0 & 20,  2 & 0.070 & 0.062 &  1.95 &  6.59 \\
\isotope[272][84]{Po}$_{188}$ & 34,  2 & 30,  4 & 0.100 & 0.093 &  4.07 &  7.37 \\
\isotope[274][84]{Po}$_{190}$ & 46,  0 & 40,  6 & 0.130 & 0.123 &  1.05 &  7.76 \\
\isotope[276][84]{Po}$_{192}$ & 52,  4 & 54,  0 & 0.152 & 0.151 &  4.46 &  0.89 \\
\isotope[278][84]{Po}$_{194}$ & 60,  4 & 62,  0 & 0.173 & 0.173 &  3.90 &  0.78 \\
\isotope[280][84]{Po}$_{196}$ & 70,  0 & 62, 10 & 0.194 & 0.188 &  0.69 &  7.92 \\
\isotope[282][84]{Po}$_{198}$ & 72,  6 & 74,  2 & 0.208 & 0.207 &  4.53 &  1.95 \\
\hline
\end{tabular}

\end{table}

\setcounter{table}{1} 
\begin{table}

\begin{tabular}{ r   r  r r c c r  }
\hline
nucleus  & hw & nhw & $\beta_{hw}$ & $\beta_{nhw}$ & $\gamma_{hw}$ & $\gamma_{nhw}$  \\ 
        
\hline

 \isotope[170][86]{Rn}$_{84}$ & 26,  2 & 22,  4 & 0.123 & 0.111 &  5.21 &  9.64 \\
 \isotope[172][86]{Rn}$_{86}$ & 32,  4 & 28,  6 & 0.153 & 0.142 &  6.95 & 10.57 \\
 \isotope[174][86]{Rn}$_{88}$ & 40,  2 & 34,  8 & 0.182 & 0.172 &  3.50 & 11.16 \\
 \isotope[176][86]{Rn}$_{90}$ & 42,  6 & 44,  2 & 0.199 & 0.198 &  7.43 &  3.20 \\
 \isotope[178][86]{Rn}$_{92}$ & 46,  6 & 48,  2 & 0.215 & 0.214 &  6.85 &  2.95 \\
 \isotope[180][86]{Rn}$_{94}$ & 52,  2 & 44, 12 & 0.230 & 0.222 &  2.73 & 12.33 \\
 \isotope[182][86]{Rn}$_{96}$ & 50,  8 & 52,  4 & 0.235 & 0.234 &  7.99 &  4.46 \\
\hline
\end{tabular}

\end{table}

\setcounter{table}{1} 
\begin{table}

\begin{tabular}{ r   r  r r c c r  }
\hline
nucleus  & hw & nhw & $\beta_{hw}$ & $\beta_{nhw}$ & $\gamma_{hw}$ & $\gamma_{nhw}$  \\ 
        
\hline

\isotope[272][86]{Rn}$_{186}$ & 30,  2 & 26,  4 & 0.089 & 0.082 &  4.57 &  8.35 \\
\isotope[274][86]{Rn}$_{188}$ & 40,  4 & 36,  6 & 0.120 & 0.112 &  5.69 &  8.51 \\
\isotope[276][86]{Rn}$_{190}$ & 52,  2 & 46,  8 & 0.149 & 0.142 &  2.73 &  8.61 \\
\isotope[278][86]{Rn}$_{192}$ & 58,  6 & 60,  2 & 0.171 & 0.170 &  5.54 &  2.38 \\
\isotope[280][86]{Rn}$_{194}$ & 66,  6 & 68,  2 & 0.192 & 0.191 &  4.92 &  2.11 \\
\isotope[282][86]{Rn}$_{196}$ & 76,  2 & 68, 12 & 0.213 & 0.207 &  1.90 &  8.48 \\
\isotope[284][86]{Rn}$_{198}$ & 78,  8 & 80,  4 & 0.227 & 0.226 &  5.33 &  2.97 \\
\hline
\end{tabular}

\end{table}

\setcounter{table}{1} 
\begin{table}

\begin{tabular}{ r   r  r r c c r  }
\hline
nucleus  & hw & nhw & $\beta_{hw}$ & $\beta_{nhw}$ & $\gamma_{hw}$ & $\gamma_{nhw}$  \\ 
        
\hline

 \isotope[172][88]{Ra}$_{84}$ & 34,  0 & 28,  6 & 0.152 & 0.142 &  1.40 & 10.57 \\
 \isotope[174][88]{Ra}$_{86}$ & 40,  2 & 34,  8 & 0.182 & 0.172 &  3.50 & 11.16 \\
 \isotope[176][88]{Ra}$_{88}$ & 48,  0 & 42,  6 & 0.210 & 0.199 &  1.00 &  7.43 \\
 \isotope[178][88]{Ra}$_{90}$ & 50,  4 & 52,  0 & 0.227 & 0.226 &  4.63 &  0.93 \\
 \isotope[180][88]{Ra}$_{92}$ & 54,  4 & 56,  0 & 0.243 & 0.242 &  4.31 &  0.86 \\
 \isotope[182][88]{Ra}$_{94}$ & 60,  0 & 52, 10 & 0.258 & 0.249 &  0.81 &  9.25 \\
 \isotope[184][88]{Ra}$_{96}$ & 58,  6 & 60,  2 & 0.263 & 0.262 &  5.54 &  2.38 \\
\hline
\end{tabular}

\end{table}

\setcounter{table}{1} 
\begin{table}

\begin{tabular}{ r   r  r r c c r  }
\hline
nucleus  & hw & nhw & $\beta_{hw}$ & $\beta_{nhw}$ & $\gamma_{hw}$ & $\gamma_{nhw}$  \\ 
        
\hline

\isotope[274][88]{Ra}$_{186}$ & 38,  0 & 32,  6 & 0.108 & 0.101 &  1.26 &  9.43 \\
\isotope[276][88]{Ra}$_{188}$ & 48,  2 & 42,  8 & 0.138 & 0.131 &  2.95 &  9.32 \\
\isotope[278][88]{Ra}$_{190}$ & 60,  0 & 54,  6 & 0.167 & 0.160 &  0.81 &  5.92 \\
\isotope[280][88]{Ra}$_{192}$ & 66,  4 & 68,  0 & 0.189 & 0.189 &  3.57 &  0.71 \\
\isotope[282][88]{Ra}$_{194}$ & 74,  4 & 76,  0 & 0.210 & 0.210 &  3.20 &  0.64 \\
\isotope[284][88]{Ra}$_{196}$ & 84,  0 & 76, 10 & 0.231 & 0.224 &  0.58 &  6.59 \\
\isotope[286][88]{Ra}$_{198}$ & 86,  6 & 88,  2 & 0.244 & 0.244 &  3.83 &  1.64 \\
\hline
\end{tabular}

\end{table}

\begin{table}

\begin{tabular}{ r   r  r r c c r  }
\hline
nucleus  & hw & nhw & $\beta_{hw}$ & $\beta_{nhw}$ & $\gamma_{hw}$ & $\gamma_{nhw}$  \\ 
        
\hline

 \isotope[174][90]{Th}$_{84}$ & 36,  4 & 38,  0 & 0.169 & 0.168 &  6.26 &  1.26 \\
 \isotope[176][90]{Th}$_{86}$ & 42,  6 & 44,  2 & 0.199 & 0.198 &  7.43 &  3.20 \\
 \isotope[178][90]{Th}$_{88}$ & 50,  4 & 52,  0 & 0.227 & 0.226 &  4.63 &  0.93 \\
 \isotope[180][90]{Th}$_{90}$ & 52,  8 & 54,  4 & 0.245 & 0.243 &  7.72 &  4.31 \\
 \isotope[182][90]{Th}$_{92}$ & 56,  8 & 58,  4 & 0.260 & 0.259 &  7.22 &  4.03 \\
 \isotope[184][90]{Th}$_{94}$ & 62,  4 & 64,  0 & 0.275 & 0.274 &  3.78 &  0.76 \\
 \isotope[186][90]{Th}$_{96}$ & 60, 10 & 62,  6 & 0.280 & 0.278 &  8.15 &  5.21 \\
\hline
\end{tabular}

\end{table}

\begin{table}

\begin{tabular}{ r   r  r r c c r  }
\hline
nucleus  & hw & nhw & $\beta_{hw}$ & $\beta_{nhw}$ & $\gamma_{hw}$ & $\gamma_{nhw}$  \\ 
        
\hline

\isotope[276][90]{Th}$_{186}$ & 40,  4 & 42,  0 & 0.119 & 0.119 &  5.69 &  1.14 \\
\isotope[278][90]{Th}$_{188}$ & 50,  6 & 52,  2 & 0.149 & 0.148 &  6.35 &  2.73 \\
\isotope[280][90]{Th}$_{190}$ & 62,  4 & 64,  0 & 0.178 & 0.178 &  3.78 &  0.76 \\
\isotope[282][90]{Th}$_{192}$ & 68,  8 & 70,  4 & 0.200 & 0.199 &  6.05 &  3.37 \\
\isotope[284][90]{Th}$_{194}$ & 76,  8 & 78,  4 & 0.221 & 0.220 &  5.46 &  3.04 \\
\isotope[286][90]{Th}$_{196}$ & 86,  4 & 88,  0 & 0.241 & 0.241 &  2.77 &  0.55 \\
\isotope[288][90]{Th}$_{198}$ & 88, 10 & 90,  6 & 0.255 & 0.255 &  5.76 &  3.67 \\
\hline
\end{tabular}

\end{table}

\begin{table}

\begin{tabular}{ r   r  r r c c r  }
\hline
nucleus  & hw & nhw & $\beta_{hw}$ & $\beta_{nhw}$ & $\gamma_{hw}$ & $\gamma_{nhw}$  \\ 
        
\hline

 \isotope[176][92]{U}$_{84}$ & 40,  4 & 42,  0 & 0.186 & 0.185 &  5.69 &  1.14 \\
 \isotope[178][92]{U}$_{86}$ & 46,  6 & 48,  2 & 0.215 & 0.214 &  6.85 &  2.95 \\
 \isotope[180][92]{U}$_{88}$ & 54,  4 & 56,  0 & 0.243 & 0.242 &  4.31 &  0.86 \\
 \isotope[182][92]{U}$_{90}$ & 56,  8 & 58,  4 & 0.260 & 0.259 &  7.22 &  4.03 \\
 \isotope[184][92]{U}$_{92}$ & 60,  8 & 62,  4 & 0.276 & 0.275 &  6.79 &  3.78 \\
 \isotope[186][92]{U}$_{94}$ & 66,  4 & 68,  0 & 0.290 & 0.290 &  3.57 &  0.71 \\
 \isotope[188][92]{U}$_{96}$ & 64, 10 & 66,  6 & 0.296 & 0.294 &  7.70 &  4.92 \\
\hline
\end{tabular}

\end{table}

\begin{table}

\begin{tabular}{ r   r  r r c c r  }
\hline
nucleus  & hw & nhw & $\beta_{hw}$ & $\beta_{nhw}$ & $\gamma_{hw}$ & $\gamma_{nhw}$  \\ 
        
\hline

\isotope[278][92]{U}$_{186}$ & 44,  4 & 46,  0 & 0.130 & 0.129 &  5.21 &  1.05 \\
\isotope[280][92]{U}$_{188}$ & 54,  6 & 56,  2 & 0.160 & 0.159 &  5.92 &  2.54 \\
\isotope[282][92]{U}$_{190}$ & 66,  4 & 68,  0 & 0.189 & 0.188 &  3.57 &  0.71 \\
\isotope[284][92]{U}$_{192}$ & 72,  8 & 74,  4 & 0.210 & 0.210 &  5.74 &  3.20 \\
\isotope[286][92]{U}$_{194}$ & 80,  8 & 82,  4 & 0.231 & 0.231 &  5.21 &  2.90 \\
\isotope[288][92]{U}$_{196}$ & 90,  4 & 92,  0 & 0.252 & 0.251 &  2.65 &  0.53 \\
\isotope[290][92]{U}$_{198}$ & 92, 10 & 94,  6 & 0.265 & 0.265 &  5.52 &  3.52 \\
\hline
\end{tabular}

\end{table}

\begin{table}

\begin{tabular}{ r   r  r r c c r  }
\hline
nucleus  & hw & nhw & $\beta_{hw}$ & $\beta_{nhw}$ & $\gamma_{hw}$ & $\gamma_{nhw}$  \\ 
        
\hline

 \isotope[178][94]{Pu}$_{84}$ & 46,  0 & 38, 10 & 0.201 & 0.193 &  1.05 & 12.08 \\
 \isotope[180][94]{Pu}$_{86}$ & 52,  2 & 44, 12 & 0.230 & 0.222 &  2.73 & 12.33 \\
 \isotope[182][94]{Pu}$_{88}$ & 60,  0 & 52, 10 & 0.258 & 0.249 &  0.81 &  9.25 \\
 \isotope[184][94]{Pu}$_{90}$ & 62,  4 & 64,  0 & 0.275 & 0.274 &  3.78 &  0.76 \\
 \isotope[186][94]{Pu}$_{92}$ & 66,  4 & 68,  0 & 0.290 & 0.290 &  3.57 &  0.71 \\
 \isotope[188][94]{Pu}$_{94}$ & 72,  0 & 64, 10 & 0.305 & 0.296 &  0.68 &  7.70 \\
 \isotope[190][94]{Pu}$_{96}$ & 70,  6 & 72,  2 & 0.310 & 0.309 &  4.65 &  2.00 \\
\hline
\end{tabular}

\end{table}

\begin{table}

\begin{tabular}{ r   r  r r c c r  }
\hline
nucleus  & hw & nhw & $\beta_{hw}$ & $\beta_{nhw}$ & $\gamma_{hw}$ & $\gamma_{nhw}$  \\ 
        
\hline

\isotope[280][94]{Pu}$_{186}$ & 50,  0 & 42, 10 & 0.140 & 0.134 &  0.96 & 11.11 \\
\isotope[282][94]{Pu}$_{188}$ & 60,  2 & 52, 12 & 0.169 & 0.164 &  2.38 & 10.71 \\
\isotope[284][94]{Pu}$_{190}$ & 72,  0 & 64, 10 & 0.199 & 0.192 &  0.68 &  7.70 \\
\isotope[286][94]{Pu}$_{192}$ & 78,  4 & 80,  0 & 0.220 & 0.220 &  3.04 &  0.61 \\
\isotope[288][94]{Pu}$_{194}$ & 86,  4 & 88,  0 & 0.241 & 0.241 &  2.77 &  0.55 \\
\isotope[290][94]{Pu}$_{196}$ & 96,  0 & 88, 10 & 0.262 & 0.255 &  0.51 &  5.76 \\
\isotope[292][94]{Pu}$_{198}$ & 98,  6 &100,  2 & 0.275 & 0.274 &  3.39 &  1.45 \\
\hline
\end{tabular}

\end{table}

\begin{table}

\begin{tabular}{ r   r  r r c c r  }
\hline
nucleus  & hw & nhw & $\beta_{hw}$ & $\beta_{nhw}$ & $\gamma_{hw}$ & $\gamma_{nhw}$  \\ 
        
\hline

 \isotope[180][96]{Cm}$_{84}$ & 44,  6 & 46,  2 & 0.206 & 0.205 &  7.13 &  3.07 \\
 \isotope[182][96]{Cm}$_{86}$ & 50,  8 & 52,  4 & 0.235 & 0.234 &  7.99 &  4.46 \\
 \isotope[184][96]{Cm}$_{88}$ & 58,  6 & 60,  2 & 0.263 & 0.262 &  5.54 &  2.38 \\
 \isotope[186][96]{Cm}$_{90}$ & 60, 10 & 62,  6 & 0.280 & 0.278 &  8.15 &  5.21 \\
 \isotope[188][96]{Cm}$_{92}$ & 64, 10 & 66,  6 & 0.296 & 0.294 &  7.70 &  4.92 \\
 \isotope[190][96]{Cm}$_{94}$ & 70,  6 & 72,  2 & 0.310 & 0.309 &  4.65 &  2.00 \\
 \isotope[192][96]{Cm}$_{96}$ & 68, 12 & 70,  8 & 0.315 & 0.313 &  8.48 &  5.89 \\
\hline
\end{tabular}

\end{table}

\begin{table}

\begin{tabular}{ r   r  r r c c r  }
\hline
nucleus  & hw & nhw & $\beta_{hw}$ & $\beta_{nhw}$ & $\gamma_{hw}$ & $\gamma_{nhw}$  \\ 
        
\hline

\isotope[282][96]{Cm}$_{186}$ & 48,  6 & 50,  2 & 0.143 & 0.142 &  6.59 &  2.83 \\
\isotope[284][96]{Cm}$_{188}$ & 58,  8 & 60,  4 & 0.173 & 0.172 &  7.00 &  3.90 \\
\isotope[286][96]{Cm}$_{190}$ & 70,  6 & 72,  2 & 0.201 & 0.201 &  4.65 &  2.00 \\
\isotope[288][96]{Cm}$_{192}$ & 76, 10 & 78,  6 & 0.223 & 0.222 &  6.59 &  4.20 \\
\isotope[290][96]{Cm}$_{194}$ & 84, 10 & 86,  6 & 0.244 & 0.243 &  6.01 &  3.83 \\
\isotope[292][96]{Cm}$_{196}$ & 94,  6 & 96,  2 & 0.264 & 0.264 &  3.52 &  1.51 \\
\isotope[294][96]{Cm}$_{198}$ & 96, 12 & 98,  8 & 0.278 & 0.277 &  6.21 &  4.31 \\
\hline
\end{tabular}

\end{table}

\begin{table}

\begin{tabular}{ r   r  r r c c r  }
\hline
nucleus  & hw & nhw & $\beta_{hw}$ & $\beta_{nhw}$ & $\gamma_{hw}$ & $\gamma_{nhw}$  \\ 
        
\hline

 \isotope[182][98]{Cf}$_{84}$ & 44,  8 & 46,  4 & 0.211 & 0.209 &  8.95 &  5.00 \\
 \isotope[184][98]{Cf}$_{86}$ & 50, 10 & 52,  6 & 0.240 & 0.238 &  9.57 &  6.12 \\
 \isotope[186][98]{Cf}$_{88}$ & 58,  8 & 60,  4 & 0.267 & 0.265 &  7.00 &  3.90 \\
 \isotope[188][98]{Cf}$_{90}$ & 60, 12 & 62,  8 & 0.284 & 0.282 &  9.47 &  6.59 \\
 \isotope[190][98]{Cf}$_{92}$ & 64, 12 & 66,  8 & 0.300 & 0.298 &  8.95 &  6.22 \\
 \isotope[192][98]{Cf}$_{94}$ & 70,  8 & 72,  4 & 0.313 & 0.312 &  5.89 &  3.28 \\
 \isotope[194][98]{Cf}$_{96}$ & 68, 14 & 70, 10 & 0.319 & 0.317 &  9.64 &  7.10 \\
\hline
\end{tabular}

\end{table}

\begin{table}

\begin{tabular}{ r   r  r r c c r  }
\hline
nucleus  & hw & nhw & $\beta_{hw}$ & $\beta_{nhw}$ & $\gamma_{hw}$ & $\gamma_{nhw}$  \\ 
        
\hline

\isotope[284][98]{Cf}$_{186}$ & 48,  8 & 50,  4 & 0.146 & 0.145 &  8.29 &  4.63 \\
\isotope[286][98]{Cf}$_{188}$ & 58, 10 & 60,  6 & 0.176 & 0.175 &  8.40 &  5.37 \\
\isotope[288][98]{Cf}$_{190}$ & 70,  8 & 72,  4 & 0.204 & 0.203 &  5.89 &  3.28 \\
\isotope[290][98]{Cf}$_{192}$ & 76, 12 & 78,  8 & 0.226 & 0.225 &  7.68 &  5.33 \\
\isotope[292][98]{Cf}$_{194}$ & 84, 12 & 86,  8 & 0.247 & 0.246 &  7.02 &  4.87 \\
\isotope[294][98]{Cf}$_{196}$ & 94,  8 & 96,  4 & 0.266 & 0.266 &  4.48 &  2.49 \\
\isotope[296][98]{Cf}$_{198}$ & 96, 14 & 98, 10 & 0.281 & 0.280 &  7.09 &  5.21 \\
\hline
\end{tabular}

\end{table}

\begin{table}

\begin{tabular}{ r   r  r r c c r  }
\hline
nucleus  & hw & nhw & $\beta_{hw}$ & $\beta_{nhw}$ & $\gamma_{hw}$ & $\gamma_{nhw}$  \\ 
        
\hline

 \isotope[184][100]{Fm}$_{84}$ & 46,  6 & 48,  2 & 0.213 & 0.212 &  6.85 &  2.95 \\
 \isotope[186][100]{Fm}$_{86}$ & 52,  8 & 54,  4 & 0.242 & 0.240 &  7.72 &  4.31 \\
 \isotope[188][100]{Fm}$_{88}$ & 60,  6 & 62,  2 & 0.269 & 0.268 &  5.37 &  2.31 \\
 \isotope[190][100]{Fm}$_{90}$ & 62, 10 & 64,  6 & 0.286 & 0.285 &  7.92 &  5.06 \\
 \isotope[192][100]{Fm}$_{92}$ & 66, 10 & 68,  6 & 0.302 & 0.300 &  7.49 &  4.78 \\
 \isotope[194][100]{Fm}$_{94}$ & 72,  6 & 74,  2 & 0.316 & 0.315 &  4.53 &  1.95 \\
 \isotope[196][100]{Fm}$_{96}$ & 70, 12 & 72,  8 & 0.321 & 0.319 &  8.27 &  5.74 \\
\hline
\end{tabular}

\end{table}

\begin{table}

\begin{tabular}{ r   r  r r c c r  }
\hline
nucleus  & hw & nhw & $\beta_{hw}$ & $\beta_{nhw}$ & $\gamma_{hw}$ & $\gamma_{nhw}$  \\ 
        
\hline

\isotope[286][100]{Fm}$_{186}$ & 50,  6 & 52,  2 & 0.148 & 0.147 &  6.35 &  2.73 \\
\isotope[288][100]{Fm}$_{188}$ & 60,  8 & 62,  4 & 0.177 & 0.176 &  6,79 &  3.78 \\
\isotope[290][100]{Fm}$_{190}$ & 72,  6 & 74,  2 & 0.206 & 0.205 &  4.53 &  1.95 \\
\isotope[292][100]{Fm}$_{192}$ & 78, 10 & 80,  6 & 0.228 & 0.227 &  6.43 &  4.10 \\
\isotope[294][100]{Fm}$_{194}$ & 86, 10 & 88,  6 & 0.248 & 0.248 &  5.88 &  3.75 \\
\isotope[296][100]{Fm}$_{196}$ & 96,  6 & 98,  2 & 0.268 & 0.268 &  3.45 &  1.48 \\
\isotope[298][100]{Fm}$_{198}$ & 98, 12 &100,  8 & 0.282 & 0.281 &  6.09 &  4.23 \\
\hline
\end{tabular}

\end{table}

\begin{table}

\begin{tabular}{ r   r  r r c c r  }
\hline
nucleus  & hw & nhw & $\beta_{hw}$ & $\beta_{nhw}$ & $\gamma_{hw}$ & $\gamma_{nhw}$  \\ 
        
\hline

 \isotope[186][102]{No}$_{84}$ & 50,  0 & 40, 14 & 0.215 & 0.209 &  0.96 & 14.99 \\
 \isotope[188][102]{No}$_{86}$ & 56,  2 & 46, 16 & 0.243 & 0.239 &  2.54 & 14.86 \\
 \isotope[190][102]{No}$_{88}$ & 64,  0 & 54, 14 & 0.271 & 0.264 &  0.76 & 11.74 \\
 \isotope[192][102]{No}$_{90}$ & 66,  4 & 68,  0 & 0.287 & 0.287 &  3.57 &  0.71 \\
 \isotope[194][102]{No}$_{92}$ & 70,  4 & 72,  0 & 0.303 & 0.302 &  3.37 &  0.68 \\
 \isotope[196][102]{No}$_{94}$ & 76,  0 & 66, 14 & 0.318 & 0.310 &  0.64 &  9.89 \\
 \isotope[198][102]{No}$_{96}$ & 74,  6 & 76,  2 & 0.322 & 0.321 &  4.42 &  1.90 \\
\hline
\end{tabular}

\end{table}

\begin{table}

\begin{tabular}{ r   r  r r c c r  }
\hline
nucleus  & hw & nhw & $\beta_{hw}$ & $\beta_{nhw}$ & $\gamma_{hw}$ & $\gamma_{nhw}$  \\ 
        
\hline

\isotope[288][102]{No}$_{186}$ & 54,  0 & 44, 14 & 0.149 & 0.145 &  0.89 & 13.90 \\
\isotope[290][102]{No}$_{188}$ & 64,  2 & 54, 16 & 0.179 & 0.175 &  2.24 & 13.05 \\
\isotope[292][102]{No}$_{190}$ & 76,  0 & 66, 14 & 0.207 & 0.202 &  0.64 &  9.89 \\
\isotope[294][102]{No}$_{192}$ & 82,  4 & 84,  0 & 0.229 & 0.228 &  2.90 &  0.58 \\
\isotope[296][102]{No}$_{194}$ & 90,  4 & 92,  0 & 0.249 & 0.249 &  2.65 &  0.53 \\
\isotope[298][102]{No}$_{196}$ &100,  0 & 90, 14 & 0.270 & 0.264 &  0.49 &  7.51 \\
\isotope[300][102]{No}$_{198}$ &102,  6 &104,  2 & 0.283 & 0.283 &  3.26 &  1.40 \\
\hline
\end{tabular}

\end{table}

\begin{table}

\begin{tabular}{ r   r  r r c c r  }
\hline
nucleus  & hw & nhw & $\beta_{hw}$ & $\beta_{nhw}$ & $\gamma_{hw}$ & $\gamma_{nhw}$  \\ 
        
\hline
 \isotope[188][104]{Rf}$_{84}$ & 44,  8 & 46,  4 & 0.208 & 0.206 &  8.95 &  5.00 \\
 \isotope[190][104]{Rf}$_{86}$ & 50, 10 & 52,  6 & 0.237 & 0.235 &  9.57 &  6.12 \\
 \isotope[192][104]{Rf}$_{88}$ & 58,  8 & 60,  4 & 0.264 & 0.263 &  7.00 &  3.90 \\
 \isotope[194][104]{Rf}$_{90}$ & 60, 12 & 62,  8 & 0.281 & 0.279 &  9.47 &  6.59 \\
 \isotope[196][104]{Rf}$_{92}$ & 64, 12 & 66,  8 & 0.297 & 0.295 &  8.95 &  6.22 \\
 \isotope[198][104]{Rf}$_{94}$ & 70,  8 & 72,  4 & 0.310 & 0.309 &  5.89 &  3.28 \\
 \isotope[200][104]{Rf}$_{96}$ & 68, 14 & 70, 10 & 0.316 & 0.314 &  9.64 &  7.10 \\
\hline
\end{tabular}

\end{table}

\begin{table}

\begin{tabular}{ r   r  r r c c r  }
\hline
nucleus  & hw & nhw & $\beta_{hw}$ & $\beta_{nhw}$ & $\gamma_{hw}$ & $\gamma_{nhw}$  \\ 
        
\hline

\isotope[290][104]{Rf}$_{186}$ & 48,  8 & 50,  4 & 0.145 & 0.144 &  8.29 &  4.63 \\
\isotope[292][104]{Rf}$_{188}$ & 58, 10 & 60,  6 & 0.174 & 0.173 &  8.40 &  5.37 \\
\isotope[294][104]{Rf}$_{190}$ & 70,  8 & 72,  4 & 0.203 & 0.202 &  5.89 &  3.28 \\
\isotope[296][104]{Rf}$_{192}$ & 76, 12 & 78,  8 & 0.224 & 0.223 &  7.68 &  5.33 \\
\isotope[298][104]{Rf}$_{194}$ & 84, 12 & 86,  8 & 0.245 & 0.244 &  7.02 &  4.87 \\
\isotope[300][104]{Rf}$_{196}$ & 94,  8 & 96,  4 & 0.265 & 0.264 &  4.48 &  2.49 \\
\isotope[302][104]{Rf}$_{198}$ & 96, 14 & 98, 10 & 0.279 & 0.278 &  7.09 &  5.21 \\
\hline
\end{tabular}

\end{table}

\begin{table}

\begin{tabular}{ r   r  r r c c r  }
\hline
nucleus  & hw & nhw & $\beta_{hw}$ & $\beta_{nhw}$ & $\gamma_{hw}$ & $\gamma_{nhw}$  \\ 
        
\hline

 \isotope[190][106]{Sg}$_{84}$ & 40, 12 & 42,  8 & 0.202 & 0.199 & 13.33 &  9.32 \\
 \isotope[192][106]{Sg}$_{86}$ & 46, 14 & 48, 10 & 0.231 & 0.228 & 13.41 &  9.92 \\
 \isotope[194][106]{Sg}$_{88}$ & 54, 12 & 56,  8 & 0.257 & 0.255 & 10.37 &  7.22 \\
 \isotope[196][106]{Sg}$_{90}$ & 56, 16 & 58, 12 & 0.275 & 0.272 & 12.67 &  9.75 \\
 \isotope[198][106]{Sg}$_{92}$ & 60, 16 & 62, 12 & 0.290 & 0.288 & 11.96 &  9.20 \\
 \isotope[200][106]{Sg}$_{94}$ & 66, 12 & 68,  8 & 0.303 & 0.301 &  8.71 &  6.05 \\
 \isotope[202][106]{Sg}$_{96}$ & 64, 18 & 66, 14 & 0.310 & 0.307 & 12.45 &  9.89 \\
\hline
\end{tabular}

\end{table}

\begin{table}

\begin{tabular}{ r   r  r r c c r  }
\hline
nucleus  & hw & nhw & $\beta_{hw}$ & $\beta_{nhw}$ & $\gamma_{hw}$ & $\gamma_{nhw}$  \\ 
        
\hline

\isotope[292][106]{Sg}$_{186}$ & 44, 12 & 46,  8 & 0.141 & 0.139 & 12.33 &  8.61 \\
\isotope[294][106]{Sg}$_{188}$ & 54, 14 & 56, 10 & 0.170 & 0.169 & 11.74 &  8.67 \\
\isotope[296][106]{Sg}$_{190}$ & 66, 12 & 68,  8 & 0.198 & 0.197 &  8.71 &  6.05 \\
\isotope[298][106]{Sg}$_{192}$ & 72, 16 & 74, 12 & 0.220 & 0.219 & 10.24 &  7.87 \\
\isotope[300][106]{Sg}$_{194}$ & 80, 16 & 82, 12 & 0.241 & 0.239 &  9.34 &  7.17 \\
\isotope[302][106]{Sg}$_{196}$ & 90, 12 & 92,  8 & 0.260 & 0.259 &  6.59 &  4.57 \\
\isotope[304][106]{Sg}$_{198}$ & 92, 18 & 94, 14 & 0.274 & 0.273 &  9.12 &  7.22 \\
\hline
\end{tabular}

\end{table}

\begin{table}

\begin{tabular}{ r   r  r r c c r  }
\hline
nucleus  & hw & nhw & $\beta_{hw}$ & $\beta_{nhw}$ & $\gamma_{hw}$ & $\gamma_{nhw}$  \\ 
        
\hline

 \isotope[192][108]{Hs}$_{84}$ & 38, 12 & 40,  8 & 0.193 & 0.191 & 13.90 &  9.72 \\
 \isotope[194][108]{Hs}$_{86}$ & 44, 14 & 46, 10 & 0.222 & 0.219 & 13.90 & 10.28 \\
 \isotope[196][108]{Hs}$_{88}$ & 52, 12 & 54,  8 & 0.248 & 0.246 & 10.71 &  7.46 \\
 \isotope[198][108]{Hs}$_{90}$ & 54, 16 & 56, 12 & 0.266 & 0.263 & 13.05 & 10.05 \\
 \isotope[200][108]{Hs}$_{92}$ & 58, 16 & 60, 12 & 0.281 & 0.279 & 12.30 &  9.47 \\
 \isotope[202][108]{Hs}$_{94}$ & 64, 12 & 66,  8 & 0.294 & 0.292 &  8.95 &  6.22 \\
 \isotope[204][108]{Hs}$_{96}$ & 62, 18 & 64, 14 & 0.301 & 0.298 & 12.79 & 10.16 \\
\hline
\end{tabular}

\end{table}

\begin{table}

\begin{tabular}{ r   r  r r c c r  }
\hline
nucleus  & hw & nhw & $\beta_{hw}$ & $\beta_{nhw}$ & $\gamma_{hw}$ & $\gamma_{nhw}$  \\ 
        
\hline

\isotope[294][108]{Hs}$_{186}$ & 42, 12 & 44,  8 & 0.136 & 0.134 & 12.81 &  8.95 \\
\isotope[296][108]{Hs}$_{188}$ & 52, 14 & 54, 10 & 0.165 & 0.163 & 12.12 &  8.95 \\
\isotope[298][108]{Hs}$_{190}$ & 64, 12 & 66,  8 & 0.192 & 0.191 &  8.95 &  6.22 \\
\isotope[300][108]{Hs}$_{192}$ & 70, 16 & 72, 12 & 0.214 & 0.213 & 10.49 &  8.06 \\
\isotope[302][108]{Hs}$_{194}$ & 78, 16 & 80, 12 & 0.235 & 0.233 &  9.55 &  7.33 \\
\isotope[304][108]{Hs}$_{196}$ & 88, 12 & 90,  8 & 0.254 & 0.253 &  6.72 &  4.67 \\
\isotope[306][108]{Hs}$_{198}$ & 90, 18 & 92, 14 & 0.268 & 0.267 &  9.30 &  7.37 \\
\hline
\end{tabular}

\end{table}

\begin{table}

\begin{tabular}{ r   r  r r c c r  }
\hline
nucleus  & hw & nhw & $\beta_{hw}$ & $\beta_{nhw}$ & $\gamma_{hw}$ & $\gamma_{nhw}$  \\ 
        
\hline

 \isotope[194][110]{Ds}$_{84}$ & 38,  8 & 40,  4 & 0.182 & 0.180 & 10.16 &  5.69 \\
 \isotope[196][110]{Ds}$_{86}$ & 44, 10 & 46,  6 & 0.211 & 0.208 & 10.68 &  6.85 \\
 \isotope[198][110]{Ds}$_{88}$ & 52,  8 & 54,  4 & 0.237 & 0.236 &  7.72 &  4.31 \\
 \isotope[200][110]{Ds}$_{90}$ & 54, 12 & 56,  8 & 0.255 & 0.252 & 10.37 &  7.22 \\
 \isotope[202][110]{Ds}$_{92}$ & 58, 12 & 60,  8 & 0.270 & 0.268 &  9.75 &  6.79 \\
 \isotope[204][110]{Ds}$_{94}$ & 64,  8 & 66,  4 & 0.283 & 0.282 &  6.40 &  3.57 \\
 \isotope[206][110]{Ds}$_{96}$ & 62, 14 & 64, 10 & 0.289 & 0.287 & 10.44 &  7.70 \\
\hline
\end{tabular}

\end{table}

\begin{table}

\begin{tabular}{ r   r  r r c c r  }
\hline
nucleus  & hw & nhw & $\beta_{hw}$ & $\beta_{nhw}$ & $\gamma_{hw}$ & $\gamma_{nhw}$  \\ 
        
\hline

\isotope[296][110]{Ds}$_{186}$ & 42,  8 & 44,  4 & 0.128 & 0.127 &  9.32 &  5.21 \\
\isotope[298][110]{Ds}$_{188}$ & 52, 10 & 54,  6 & 0.158 & 0.156 &  9.25 &  5.92 \\
\isotope[300][110]{Ds}$_{190}$ & 64,  8 & 66,  4 & 0.185 & 0.185 &  6.40 &  3.57 \\
\isotope[302][110]{Ds}$_{192}$ & 70, 12 & 72,  8 & 0.207 & 0.206 &  8.27 &  5.74 \\
\isotope[304][110]{Ds}$_{194}$ & 78, 12 & 80,  8 & 0.228 & 0.227 &  7.50 &  5.21 \\
\isotope[306][110]{Ds}$_{196}$ & 88,  8 & 90,  4 & 0.247 & 0.247 &  4.77 &  2.65 \\
\isotope[308][110]{Ds}$_{198}$ & 90, 14 & 92, 10 & 0.261 & 0.260 &  7.51 &  5.52 \\
\hline
\end{tabular}

\end{table}

\begin{table}

\begin{tabular}{ r   r  r r c c r  }
\hline
nucleus  & hw & nhw & $\beta_{hw}$ & $\beta_{nhw}$ & $\gamma_{hw}$ & $\gamma_{nhw}$  \\ 
        
\hline

 \isotope[196][112]{Cn}$_{84}$ & 40,  0 & 28, 18 & 0.170 & 0.172 &  1.20 & 23.14 \\
 \isotope[198][112]{Cn}$_{86}$ & 46,  2 & 34, 20 & 0.198 & 0.200 &  3.07 & 21.79 \\
 \isotope[200][112]{Cn}$_{88}$ & 54,  0 & 42, 18 & 0.226 & 0.224 &  0.89 & 17.40 \\
 \isotope[202][112]{Cn}$_{90}$ & 56,  4 & 58,  0 & 0.242 & 0.241 &  4.16 &  0.83 \\
 \isotope[204][112]{Cn}$_{92}$ & 60,  4 & 62,  0 & 0.257 & 0.257 &  3.90 &  0.78 \\
 \isotope[206][112]{Cn}$_{94}$ & 66,  0 & 54, 18 & 0.272 & 0.268 &  0.74 & 14.31 \\
 \isotope[208][112]{Cn}$_{96}$ & 64,  6 & 66,  2 & 0.276 & 0.275 &  5.06 &  2.17 \\
\hline
\end{tabular}

\end{table}

\begin{table}

\begin{tabular}{ r   r  r r c c r  }
\hline
nucleus  & hw & nhw & $\beta_{hw}$ & $\beta_{nhw}$ & $\gamma_{hw}$ & $\gamma_{nhw}$  \\ 
        
\hline

\isotope[298][112]{Cn}$_{186}$ & 44,  0 & 32, 18 & 0.121 & 0.121 &  1.09 & 21.16 \\
\isotope[300][112]{Cn}$_{188}$ & 54,  2 & 42, 20 & 0.150 & 0.150 &  2.63 & 18.77 \\
\isotope[302][112]{Cn}$_{190}$ & 66,  0 & 54, 18 & 0.179 & 0.176 &  0.74 & 14.31 \\
\isotope[304][112]{Cn}$_{192}$ & 72,  4 & 74,  0 & 0.200 & 0.199 &  3.28 &  0.66 \\
\isotope[306][112]{Cn}$_{194}$ & 80,  4 & 82,  0 & 0.220 & 0.220 &  2.97 &  0.59 \\
\isotope[308][112]{Cn}$_{196}$ & 90,  0 & 78, 18 & 0.241 & 0.237 &  0.54 & 10.53 \\
\isotope[310][112]{Cn}$_{198}$ & 92,  6 & 94,  2 & 0.254 & 0.253 &  3.60 &  1.54 \\
\hline
\end{tabular}

\end{table}

\begin{table}

\begin{tabular}{ r   r  r r c c r  }
\hline
nucleus  & hw & nhw & $\beta_{hw}$ & $\beta_{nhw}$ & $\gamma_{hw}$ & $\gamma_{nhw}$  \\ 
        
\hline

 \isotope[198][114]{Fl}$_{84}$ & 30, 10 & 32,  6 & 0.154 & 0.151 & 14.63 &  9.43 \\
 \isotope[200][114]{Fl}$_{86}$ & 36, 12 & 38,  8 & 0.183 & 0.180 & 14.51 & 10.16 \\
 \isotope[202][114]{Fl}$_{88}$ & 44, 10 & 46,  6 & 0.209 & 0.206 & 10.68 &  6.85 \\
 \isotope[204][114]{Fl}$_{90}$ & 46, 14 & 48, 10 & 0.227 & 0.224 & 13.41 &  9.92 \\
 \isotope[206][114]{Fl}$_{92}$ & 50, 14 & 52, 10 & 0.242 & 0.239 & 12.52 &  9.25 \\
 \isotope[208][114]{Fl}$_{94}$ & 56, 10 & 58,  6 & 0.254 & 0.252 &  8.67 &  5.54 \\
 \isotope[210][114]{Fl}$_{96}$ & 54, 16 & 56, 12 & 0.261 & 0.258 & 13.05 & 10.05 \\
\hline
\end{tabular}

\end{table}

\begin{table}

\begin{tabular}{ r   r  r r c c r  }
\hline
nucleus  & hw & nhw & $\beta_{hw}$ & $\beta_{nhw}$ & $\gamma_{hw}$ & $\gamma_{nhw}$  \\ 
        
\hline

\isotope[300][114]{Fl}$_{186}$ & 34, 10 & 36,  6 & 0.110 & 0.109 & 13.24 &  8.51 \\
\isotope[302][114]{Fl}$_{188}$ & 44, 12 & 46,  8 & 0.140 & 0.138 & 12.33 &  8.61 \\
\isotope[304][114]{Fl}$_{190}$ & 56, 10 & 58,  6 & 0.167 & 0.166 &  8.67 &  5.54 \\
\isotope[306][114]{Fl}$_{192}$ & 62, 14 & 64, 10 & 0.189 & 0.187 & 10.44 &  7.70 \\
\isotope[308][114]{Fl}$_{194}$ & 70, 14 & 72, 10 & 0.209 & 0.208 &  9.40 &  6.92 \\
\isotope[310][114]{Fl}$_{196}$ & 80, 10 & 82,  6 & 0.228 & 0.227 &  6.29 &  4.01 \\
\isotope[312][114]{Fl}$_{198}$ & 82, 16 & 84, 12 & 0.243 & 0.241 &  9.14 &  7.02 \\
\hline
\end{tabular}

\end{table}

\begin{table}

\begin{tabular}{ r   r  r r c c r  }
\hline
nucleus  & hw & nhw & $\beta_{hw}$ & $\beta_{nhw}$ & $\gamma_{hw}$ & $\gamma_{nhw}$  \\ 
        
\hline

 \isotope[200][116]{Lv}$_{84}$ & 22, 16 & 24, 12 & 0.142 & 0.136 & 25.05 & 19.67 \\
 \isotope[202][116]{Lv}$_{86}$ & 28, 18 & 30, 14 & 0.170 & 0.165 & 23.14 & 18.65 \\
 \isotope[204][116]{Lv}$_{88}$ & 36, 16 & 38, 12 & 0.193 & 0.190 & 17.93 & 13.90 \\
 \isotope[206][116]{Lv}$_{90}$ & 38, 20 & 40, 16 & 0.213 & 0.208 & 20.17 & 16.56 \\
 \isotope[208][116]{Lv}$_{92}$ & 42, 20 & 44, 16 & 0.227 & 0.223 & 18.77 & 15.39 \\
 \isotope[210][116]{Lv}$_{94}$ & 48, 16 & 50, 12 & 0.238 & 0.235 & 14.36 & 11.08 \\
 \isotope[212][116]{Lv}$_{96}$ & 46, 22 & 48, 18 & 0.247 & 0.243 & 18.80 & 15.71 \\
\hline
\end{tabular}

\end{table}

\begin{table}

\begin{tabular}{ r   r  r r c c r  }
\hline
nucleus  & hw & nhw & $\beta_{hw}$ & $\beta_{nhw}$ & $\gamma_{hw}$ & $\gamma_{nhw}$  \\ 
        
\hline

\isotope[302][116]{Lv}$_{186}$ & 26, 16 & 28, 12 & 0.102 & 0.099 & 22.52 & 17.60 \\
\isotope[304][116]{Lv}$_{188}$ & 36, 18 & 38, 14 & 0.130 & 0.128 & 19.49 & 15.61 \\
\isotope[306][116]{Lv}$_{190}$ & 48, 16 & 50, 12 & 0.156 & 0.154 & 14.36 & 11.08 \\
\isotope[308][116]{Lv}$_{192}$ & 54, 20 & 56, 16 & 0.179 & 0.177 & 15.52 & 12.67 \\
\isotope[310][116]{Lv}$_{194}$ & 62, 20 & 64, 16 & 0.199 & 0.197 & 13.90 & 11.33 \\
\isotope[312][116]{Lv}$_{196}$ & 72, 16 & 74, 12 & 0.217 & 0.215 & 10.24 &  7.87 \\
\isotope[314][116]{Lv}$_{198}$ & 74, 22 & 76, 18 & 0.232 & 0.230 & 12.97 & 10.77 \\
\hline
\end{tabular}

\end{table}

\begin{table}

\begin{tabular}{ r   r  r r c c r  }
\hline
nucleus  & hw & nhw & $\beta_{hw}$ & $\beta_{nhw}$ & $\gamma_{hw}$ & $\gamma_{nhw}$  \\ 
        
\hline

 \isotope[202][118]{Og}$_{84}$ & 16, 18 & 18, 14 & 0.127 & 0.120 & 31.84 & 26.11 \\
 \isotope[204][118]{Og}$_{86}$ & 22, 20 & 24, 16 & 0.154 & 0.148 & 28.50 & 23.72 \\
 \isotope[206][118]{Og}$_{88}$ & 30, 18 & 32, 14 & 0.176 & 0.171 & 22.11 & 17.78 \\
 \isotope[208][118]{Og}$_{90}$ & 32, 22 & 34, 18 & 0.196 & 0.191 & 24.11 & 20.29 \\
 \isotope[210][118]{Og}$_{92}$ & 36, 22 & 38, 18 & 0.210 & 0.205 & 22.33 & 18.74 \\
 \isotope[212][118]{Og}$_{94}$ & 42, 18 & 44, 14 & 0.220 & 0.216 & 17.40 & 13.90 \\
 \isotope[214][118]{Og}$_{96}$ & 40, 24 & 42, 20 & 0.230 & 0.225 & 22.03 & 18.77 \\
\hline
\end{tabular}

\end{table}

\begin{table}

\begin{tabular}{ r   r  r r c c r  }
\hline
nucleus  & hw & nhw & $\beta_{hw}$ & $\beta_{nhw}$ & $\gamma_{hw}$ & $\gamma_{nhw}$  \\ 
        
\hline

\isotope[304][118]{Og}$_{186}$ & 20, 18 & 22, 14 & 0.091 & 0.088 & 28.35 & 23.07 \\
\isotope[306][118]{Og}$_{188}$ & 30, 20 & 32, 16 & 0.119 & 0.116 & 23.66 & 19.53 \\
\isotope[308][118]{Og}$_{190}$ & 42, 18 & 44, 14 & 0.145 & 0.142 & 17.40 & 13.90 \\
\isotope[310][118]{Og}$_{192}$ & 48, 22 & 50, 18 & 0.167 & 0.164 & 18.22 & 15.21 \\
\isotope[312][118]{Og}$_{194}$ & 56, 22 & 58, 18 & 0.187 & 0.184 & 16.21 & 13.51 \\
\isotope[314][118]{Og}$_{196}$ & 66, 18 & 68, 14 & 0.204 & 0.203 & 12.14 &  9.64 \\
\isotope[316][118]{Og}$_{198}$ & 68, 24 & 70, 20 & 0.220 & 0.218 & 14.88 & 12.58 \\
\hline
\end{tabular}

\end{table}

\begin{table}

\begin{tabular}{ r   r  r r c c r  }
\hline
nucleus  & hw & nhw & $\beta_{hw}$ & $\beta_{nhw}$ & $\gamma_{hw}$ & $\gamma_{nhw}$  \\ 
        
\hline

 \isotope[204][120]{120}$_{84}$ & 12, 16 & 14, 12 & 0.105 & 0.098 & 34.40 & 27.64 \\
 \isotope[206][120]{120}$_{86}$ & 18, 18 & 20, 14 & 0.133 & 0.126 & 30.00 & 24.50 \\
 \isotope[208][120]{120}$_{88}$ & 26, 16 & 28, 12 & 0.154 & 0.150 & 22.52 & 17.60 \\
 \isotope[210][120]{120}$_{90}$ & 28, 20 & 30, 16 & 0.174 & 0.169 & 24.72 & 20.44 \\
 \isotope[212][120]{120}$_{92}$ & 32, 20 & 34, 16 & 0.188 & 0.183 & 22.69 & 18.70 \\
 \isotope[214][120]{120}$_{94}$ & 38, 16 & 40, 12 & 0.198 & 0.194 & 17.22 & 13.33 \\
 \isotope[216][120]{120}$_{96}$ & 36, 22 & 38, 18 & 0.208 & 0.203 & 22.33 & 18.74 \\
\hline
\end{tabular}

\end{table}

\begin{table}

\begin{tabular}{ r   r  r r c c r  }
\hline
nucleus  & hw & nhw & $\beta_{hw}$ & $\beta_{nhw}$ & $\gamma_{hw}$ & $\gamma_{nhw}$  \\ 
        
\hline

\isotope[306][120]{120}$_{186}$ & 16, 16 & 18, 12 & 0.078 & 0.073 & 30.00 & 23.82 \\
\isotope[308][120]{120}$_{188}$ & 26, 18 & 28, 14 & 0.105 & 0.102 & 24.27 & 19.59 \\
\isotope[310][120]{120}$_{190}$ & 38, 16 & 40, 12 & 0.130 & 0.128 & 17.22 & 13.33 \\
\isotope[312][120]{120}$_{192}$ & 44, 20 & 46, 16 & 0.153 & 0.150 & 18.14 & 14.86 \\
\isotope[314][120]{120}$_{194}$ & 52, 20 & 54, 16 & 0.173 & 0.170 & 15.98 & 13.05 \\
\isotope[316][120]{120}$_{196}$ & 62, 16 & 64, 12 & 0.190 & 0.189 & 11.64 &  8.95 \\
\isotope[318][120]{120}$_{198}$ & 64, 22 & 66, 18 & 0.206 & 0.204 & 14.59 & 12.14 \\
\hline
\end{tabular}

\end{table}

\begin{table}

\begin{tabular}{ r   r  r r c c r  }
\hline
nucleus  & hw & nhw & $\beta_{hw}$ & $\beta_{nhw}$ & $\gamma_{hw}$ & $\gamma_{nhw}$  \\ 
        
\hline

 \isotope[206][122]{122}$_{84}$ & 10, 10 & 12,  6 & 0.077 & 0.071 & 30.00 & 20.17 \\
 \isotope[208][122]{122}$_{86}$ & 16, 12 & 18,  8 & 0.105 & 0.099 & 25.60 & 18.35 \\
 \isotope[210][122]{122}$_{88}$ & 24, 10 & 26,  6 & 0.128 & 0.125 & 17.35 & 11.24 \\
 \isotope[212][122]{122}$_{90}$ & 26, 14 & 28, 10 & 0.147 & 0.143 & 20.63 & 15.44 \\
 \isotope[214][122]{122}$_{92}$ & 30, 14 & 32, 10 & 0.162 & 0.158 & 18.65 & 13.90 \\
 \isotope[216][122]{122}$_{94}$ & 36, 10 & 38,  6 & 0.173 & 0.170 & 12.63 &  8.12 \\
 \isotope[218][122]{122}$_{96}$ & 34, 16 & 36, 12 & 0.182 & 0.178 & 18.70 & 14.51 \\
\hline
\end{tabular}

\end{table}

\begin{table}

\begin{tabular}{ r   r  r r c c r  }
\hline
nucleus  & hw & nhw & $\beta_{hw}$ & $\beta_{nhw}$ & $\gamma_{hw}$ & $\gamma_{nhw}$  \\ 
        
\hline

\isotope[308][122]{122}$_{186}$ & 14, 10 & 16,  6 & 0.059 & 0.056 & 24.92 & 16.47 \\
\isotope[310][122]{122}$_{188}$ & 24, 12 & 26,  8 & 0.088 & 0.085 & 19.67 & 13.90 \\
\isotope[312][122]{122}$_{190}$ & 36, 10 & 38,  6 & 0.114 & 0.112 & 12.63 &  8.12 \\
\isotope[314][122]{122}$_{192}$ & 42, 14 & 44, 10 & 0.136 & 0.134 & 14.43 & 10.68 \\
\isotope[316][122]{122}$_{194}$ & 50, 14 & 52, 10 & 0.156 & 0.155 & 12.52 &  9.25 \\
\isotope[318][122]{122}$_{196}$ & 60, 10 & 62,  6 & 0.175 & 0.174 &  8.15 &  5.21 \\
\isotope[320][122]{122}$_{198}$ & 62, 16 & 64, 12 & 0.189 & 0.188 & 11.64 &  8.95 \\
\hline
\end{tabular}

\end{table}

\begin{table}

\begin{tabular}{ r   r  r r c c r  }
\hline
nucleus  & hw & nhw & $\beta_{hw}$ & $\beta_{nhw}$ & $\gamma_{hw}$ & $\gamma_{nhw}$  \\ 
        
\hline
 \isotope[208][124]{124}$_{84}$ & 10,  0 &  6,  2 & 0.046 & 0.036 &  4.31 & 17.00 \\
 \isotope[210][124]{124}$_{86}$ & 16,  2 & 12,  4 & 0.075 & 0.064 &  7.99 & 15.61 \\
 \isotope[212][124]{124}$_{88}$ & 24,  0 & 18,  6 & 0.102 & 0.093 &  1.95 & 15.08 \\
 \isotope[214][124]{124}$_{90}$ & 26,  4 & 28,  0 & 0.119 & 0.117 &  8.35 &  1.68 \\
 \isotope[216][124]{124}$_{92}$ & 30,  4 & 32,  0 & 0.134 & 0.133 &  7.37 &  1.48 \\
 \isotope[218][124]{124}$_{94}$ & 36,  0 & 28, 10 & 0.148 & 0.142 &  1.32 & 15.44 \\
 \isotope[220][124]{124}$_{96}$ & 34,  6 & 36,  2 & 0.154 & 0.152 &  8.95 &  3.86 \\
\hline
\end{tabular}

\end{table}

\begin{table}

\begin{tabular}{ r   r  r r c c r  }
\hline
nucleus  & hw & nhw & $\beta_{hw}$ & $\beta_{nhw}$ & $\gamma_{hw}$ & $\gamma_{nhw}$  \\ 
        
\hline

\isotope[310][124]{124}$_{186}$ & 14,  0 & 10,  2 & 0.041 & 0.033 &  3.20 & 11.74 \\
\isotope[312][124]{124}$_{188}$ & 24,  2 & 20,  4 & 0.070 & 0.063 &  5.60 & 10.44 \\
\isotope[314][124]{124}$_{190}$ & 36,  0 & 30,  6 & 0.098 & 0.091 &  1.32 &  9.97 \\
\isotope[316][124]{124}$_{192}$ & 42,  4 & 44,  0 & 0.119 & 0.119 &  5.44 &  1.09 \\
\isotope[318][124]{124}$_{194}$ & 50,  4 & 52,  0 & 0.140 & 0.139 &  4.63 &  0.93 \\
\isotope[320][124]{124}$_{196}$ & 60,  0 & 52, 10 & 0.160 & 0.154 &  0.81 &  9.25 \\
\isotope[322][124]{124}$_{198}$ & 62,  6 & 64,  2 & 0.173 & 0.172 &  5.21 &  2.24 \\
\hline
\end{tabular}

\end{table}